\documentclass[12pt]{article}

\usepackage[a4paper, margin=1in]{geometry}
\usepackage{amsmath, amssymb, graphicx}
\usepackage{hyperref}
\usepackage{cite}
\usepackage{booktabs}
\usepackage{adjustbox}
\usepackage{subcaption} 

\usepackage{placeins}    

\usepackage[font=footnotesize,labelfont=bf]{caption} 
\usepackage{pdflscape}
\usepackage{float}

\usepackage{authblk}
\usepackage{makecell}
\usepackage{multirow} 

\title{Recovering Structural Organization in Noisy Correlation Networks Using Financial Systems as a Testbed}

\author[1]{Imran Ansari\thanks{Corresponding author: \texttt{imranansari@iisc.ac.in}}}
\author[1]{Shashi Jain}
\author[2]{Srikanth K. Iyer}

\affil[1]{Department of Management Studies, Indian Institute of Science (IISc), Bangalore, India}

\affil[2]{Department of Mathematics, Indian Institute of Science (IISc), Bangalore, India}
\date{}

\begin{document}

\maketitle

\begin{abstract}
Empirical correlation matrices estimated from financial return time series are contaminated by substantial statistical noise arising from finite sample size, obscuring the genuine interaction structure among assets. We apply spectral theory to decompose the empirical correlation matrix into a structured component associated with eigenvalues exceeding the Marchenko–Pastur bounds and a random component consistent with statistical noise. Using daily returns from three equity markets (NIFTY 200, NIFTY 500, and S\&P 500) spanning 2010 to 2022, we show that the structured component, constructed from only 10 to 16 eigenmodes, reproduces the principal statistical properties of the full correlation matrix while eliminating the majority of noise dominated eigenmodes. We then construct financial networks from the full, structured, and random correlation matrices and characterize their core-periphery organization using a Markov chain based method. The structured network exhibits significantly stronger and more stable core periphery organization than full or random networks. This observation is supported by degree preserving randomization tests together with Kolmogorov Smirnov and Wasserstein distance analyzes, which demonstrate a clear statistical separation between the structured and random components. We further show that the structured networks exhibit pronounced scale free degree distributions in the Indian markets, whereas the random networks do not, indicating that genuine economic structure and random noise generate fundamentally distinct network topologies. As a concrete demonstration of this framework's practical relevance, we further show that portfolios constructed from peripheral assets of the denoised network achieve consistently higher risk-adjusted performance than portfolios built from unfiltered correlations or standard benchmarks, with robustness confirmed via Monte Carlo subsampling. These results establish spectral denoising as a general methodology for recovering genuine structural organization in noise-corrupted correlation-based systems, of which financial portfolio construction is one illustrative application.
\end{abstract}

\section{Introduction}
Complex systems composed of many interacting units are ubiquitous, spanning biological, technological, and social domains, and are frequently characterized through the pairwise correlations observed among their constituent variables \cite{mantegna1999hierarchical, plerou2001collective}. A central obstacle to extracting genuine interaction structure from such systems is that empirical correlation matrices, estimated from finite time series, are inherently contaminated by statistical noise. When the number of observed variables is comparable to the length of the available time series, a regime routinely encountered in high-dimensional real-world data, a substantial fraction of the eigenspectrum of the empirical correlation matrix is indistinguishable from that of a purely random matrix \cite{Laloux1999noise, Plerou1999universal}. Distinguishing the small subset of eigenmodes that encode genuine collective structure from the majority that reflect sampling noise is therefore a prerequisite for any reliable downstream analysis of network topology, community organization, or predictive modeling built upon such correlations.

Financial markets provide a particularly well-studied and data-rich setting in which to address this problem. Stock return correlations are shaped by a mixture of genuine collective dynamics, driven by common macroeconomic factors, sectoral comovement, and firm-level interdependence, and substantial estimation noise arising from finite sample size \cite{jiang2014structure}. Portfolio optimization, which relies directly on accurate estimates of the covariance structure among assets, is especially sensitive to this contamination. Classical mean-variance frameworks \cite{Markowitz1952,markowitz2000mean}, higher-moment extensions \cite{samuelson1970fundamental}, mean absolute deviation approaches \cite{konno1991mean,konno2005mean}, and downside-risk formulations \cite{nawrocki1992characteristics} all presuppose that the estimated correlation structure reflects genuine asset interdependence rather than noise, and their performance degrades accordingly when this assumption fails. This makes financial markets an ideal empirical setting for testing whether principled noise filtering of correlation matrices can recover interpretable, economically meaningful structure and, in turn, whether that structure has practical consequences for a canonical downstream task such as portfolio construction. Recent advances have explored data-driven and heuristic approaches, including reinforcement learning and evolutionary optimization techniques, to address these limitations \cite{moody2001learning,aboussalah2020continuous,yusoff2011overview,reyes2006multi}. While flexible, these methods often suffer from limited interpretability, sensitivity to hyperparameters, and substantial computational costs. This has motivated increasing interest in network-based approaches, which aim to extract stable structural information from financial markets by representing assets as interconnected systems.

A central challenge in this context is that empirical correlation matrices are inherently noisy due to finite sample effects. To mitigate this issue, prior studies have analyzed the statistical properties of empirical correlation matrices through their spectral decomposition, noting that the bulk of eigenvalues can be well approximated by the Marchenko--Pastur distribution expected for purely random correlations. Specifically, eigenvalues within the Marchenko--Pastur bounds are consistent with random noise, whereas significant deviations correspond to genuine market structure, including collective and sectoral dynamics \cite{plerou2001collective,jiang2014structure}. While prior studies~\cite{kim2005systematic, pawanesh2025exploring} often focus on isolating the dominant market mode or a small number of sectoral components, it remains unclear how the full set of statistically significant eigenmodes can be leveraged for portfolio construction, particularly in conjunction with higher-order network structures.

In parallel, network representations of financial markets have provided valuable insights into systemic risk and market organization \cite{pozzi2013spread,ansari2025novel}. Beyond local connectivity measures, recent work highlights the importance of mesoscale structures, especially core--periphery (CP) organization~\cite{borgatti2000models, ansari2025uncovering, rombach2017core} in capturing the hierarchical structure of financial systems \cite{ansari2025novel, lee2014density}. Economically, core assets tend to be strongly interconnected and highly exposed to common market factors, whereas peripheral assets exhibit weaker correlations and greater idiosyncratic behavior. This suggests that peripheral assets may offer enhanced diversification benefits, although their systematic exploitation in portfolio design remains underexplored. Against this backdrop, a key gap in the literature is the lack of an integrated framework that combines statistically grounded correlation denoising with mesoscale network structure for portfolio optimization. Existing studies typically treat these components in isolation, leaving unclear how their interaction influences the structural and economic properties of the resulting networks. 

In this work, we address these open questions by developing a unified framework that combines spectral denoising with core-periphery network analysis and systematically evaluates its consequences across four complementary dimensions. First, we construct a denoised correlation matrix by isolating the eigenmodes lying beyond the Marchenko--Pastur bounds and demonstrate that this structured component, despite using only a small fraction of the full eigenspectrum, faithfully reproduces the correlation distribution of the original data. Second, we compare the core periphery organization of networks constructed from the full, structured, and random correlation matrices, showing that the structured component yields a markedly stronger, more stable, and statistically significant core periphery structure than either alternative. Third, we characterize the scale free properties of the resulting networks and find that structured networks exhibit power law degree distributions consistent with genuine hub dominated organization, whereas random networks do not. This result reinforces that the observed topology reflects genuine market structure rather than statistical artifacts. Fourth, we perform a systematic distributional analysis of the full, structured, and random correlation matrices together with their derived network statistics, quantifying the separation between signal and noise using the Kolmogorov--Smirnov and Wasserstein metrics. Finally, to demonstrate the practical relevance of the proposed framework, we apply it to portfolio construction and show that portfolios built from peripheral assets of the structured network consistently outperform several standard benchmarks across three major equity markets. Taken together, these results establish correlation matrix denoising as more than a preprocessing step for portfolio optimization. Instead, it provides a general methodology for uncovering the genuine structural organization of complex, noise corrupted, correlation based systems.


\section{Methods}

\subsection{Data description}

This study uses three datasets from the Indian National Stock Exchange (NSE) and the U.S. stock market, namely the NIFTY~200, NIFTY~500, and S\&P~500 indices. The datasets comprise daily closing prices of all constituent stocks from January~1,~2010, to December~31,~2022. To ensure data completeness, stocks with missing observations were excluded. This filtering resulted in refined datasets consisting of 140 stocks over 3{,}208 trading days for the NIFTY~200, 312 stocks over 3{,}204 trading days for the NIFTY~500, and 425 stocks over 3{,}271 trading days for the S\&P~500 index.

The daily logarithmic return for stock \(i\) at time \(t\) is defined as
\[
r_i(t) = \ln \left (\frac{p_i(t)} { p_i(t-1)}\right),
\]
where \(p_i(t)\) denotes the closing price of stock \(i\) on day \(t\).

\begin{figure}[htbp]
    \centering
    \includegraphics[width=0.8\textwidth]{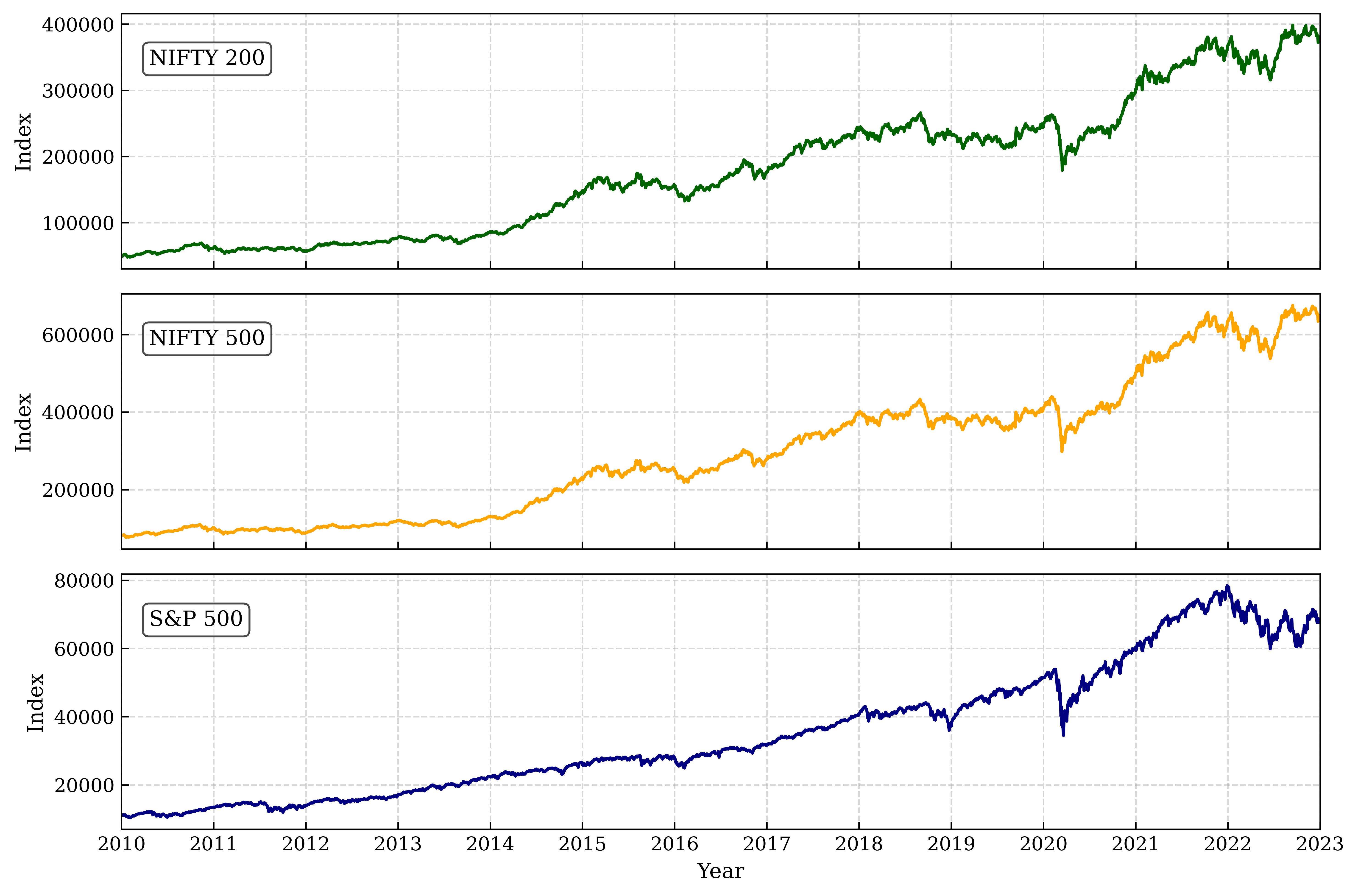}
    \caption{Time series plot of NIFTY 200, NIFTY 500 and S\&P 500 indices.}
    \label{fig:nifty}
\end{figure}

Figure~\ref{fig:nifty} shows the NIFTY~200 index, the NIFTY~500 index, and the S\&P~500 index, providing an overview of the three stock markets. This datasets includes the COVID-19 pandemic period, which captures a global crisis phase. Incorporating this period allows us to examine portfolio behavior and risk dynamics under extreme market stress conditions.


\subsection{Correlation matrix construction and spectral decomposition}

Let $r_i(t)$ denote the daily logarithmic return of the stock $i$, where $i = 1,2,\ldots,N$ and $N$ is the total number of stocks in the data set. The time index spans $t = 1,2,\ldots,T$, with $T$ denoting the total number of observations. Since different stocks exhibit varying levels of volatility ( measured by the standard deviation of their returns), we define the normalized return as
\begin{equation}
\tilde{r}_i(t) = \frac{r_i(t) - \langle r_i \rangle}{\sigma_i},
\end{equation}
where $\sigma_i = \sqrt{\langle r_i^2 \rangle - \langle r_i \rangle^2}$, is the standard deviation of $r_i(t)$, and $\langle \cdot \rangle$ denotes the time average over the observation period.

The cross-correlation matrix $C = [C_{ij}]$ is constructed from the normalized return series $\tilde{r}_i(t)$ as
\begin{equation}
C_{ij} = \left\langle \tilde{r}_i(t)\ \tilde{r}_j(t) \right\rangle
\end{equation}

The resulting matrix $C$ is real and symmetric, with diagonal elements $C_{ii}=1$, and all entries satisfying $-1 \leq C_{ij} \leq 1$.

If the $N$ return time series of length $T$ consists of independent and identically distributed Gaussian random variables with zero mean and unit variance, then the resulting random correlation matrix follows a Wishart distribution.

In the asymptotic limit $N \to \infty$ and $T \to \infty$ with the ratio 

\[
Q = \frac{T}{N} \geq 1
\]

held fixed, the probability density function of the eigenvalues $\lambda$ of the random correlation matrix converges to the Marchenko--Pastur distribution~\cite{pastur2011eigenvalue}, given by

\begin{equation}
P_{\text{rm}}(\lambda) =
\begin{cases}
\displaystyle
\frac{Q}{2\pi \lambda} 
\sqrt{(\lambda_{+} - \lambda)(\lambda - \lambda_{-})}, 
& \text{if } \lambda_{-} \leq \lambda \leq \lambda_{+}, \\[10pt]
0, & \text{otherwise},
\end{cases}
\end{equation}

where $\lambda_{-}$ and $\lambda_{+}$ denote the lower and upper bounds of the eigenvalue spectrum, defined as

\begin{equation}
\lambda_{-} = \left(1 - \frac{1}{\sqrt{Q}}\right)^2, 
\qquad 
\lambda_{+} = \left(1 + \frac{1}{\sqrt{Q}}\right)^2.
\end{equation}

The cross-correlation matrix \(C\) can be decomposed into a set of orthogonal eigenmodes as
\[
C_{ij} = \sum_{\gamma=1}^{N} C_{ij}^{\gamma},
\]
where
\[
C_{ij}^{\gamma} = \lambda_{\gamma} u_i^{\gamma} u_j^{\gamma}.
\]
Here, \(\lambda_{\gamma}\) denotes the \(\gamma\)-th eigenvalue of \(C\), and \(u_i^{\gamma}\) represents the \(i\)-th component of the corresponding eigenvector.

We decomposed the empirical cross-correlation matrix into two components—namely, the structured mode, and the random mode as follows. The \emph{structured mode} $C^{\mathrm{STR}}$ is given by $C^{\mathrm{STR}}_{ij} = \sum_{\gamma=1}^{k} C^{\gamma}_{ij}$, where \(k\) denotes the number of eigenvalues lying above the upper bound of the Marchenko--Pastur interval \([\lambda_{-}, \lambda_{+}]\), capturing the structured (non-random) correlations. The \emph{random mode} $C^{\mathrm{RND}}$ is defined as $C^{\mathrm{RND}}_{ij} = \sum_{\gamma=k+1}^{N} C^{\gamma}_{ij}$, which represents noise-driven correlations.

\subsection{Core–periphery detection algorithm}

Network science has devoted considerable effort to uncovering the core–periphery structure, a key mesoscale organization pattern observed in diverse real-world networks. Detecting core–periphery structure in networks is crucial for uncovering hierarchical relationships and understanding the organization of complex systems~\cite{borgatti2000models,boyd2006computing,rombach2017core}. 

Our study employs the iterative algorithm to extract the \textit{core–periphery profile} of the network as proposed in~\cite{rossa2013profiling}. This algorithm is grounded in Markov chain theory and provides a probabilistic measure of persistence within the core and periphery subsets. It offers a rigorous and comprehensive framework for analyzing core–periphery structures across a wide range of network types. In this formulation, the vertices of the network correspond to the states of a Markov chain, and the transition probability matrix \( (p_{ij}) \) is defined as $p_{ij} = \frac{a_{ij}}{\displaystyle\sum_{h \in V(G)} a_{ih}}$, where \( a_{ij} \) denotes the weight of the edge between vertices \( i \) and \( j \), with \( i,j,h \in V(G) \). 

Let \(S \subseteq G\) denote a subgraph of \(G\), with vertex set \(V(S) \subseteq V(G)\). The \textit{persistence probability} \(\gamma_{\strut S} \) quantifies the internal cohesiveness of \(S\) and is defined as
\[
\gamma_{\strut S}  =  \frac{\displaystyle\sum_{i,j \in V(S)} \pi_i p_{ij}}{\displaystyle\sum_{i \in V(S)} \pi_i},
\]
where \(\pi_i > 0\) is the stationary probability of being at vertex \(i\)~\cite{meyer2023matrix}, given by
\[
\pi_i = \frac{C_D(i)}{\displaystyle\sum_{j \in V(G)} C_D(j)} \quad \text{with} \quad 
C_D(i) = \sum_{h \in V(G)} a_{ih},
\]
and \(C_D(i)\) denotes the weighted degree (strength) of vertex \(i\). Using this formulation, \(\gamma_{\strut S} \) simplifies to
\begin{equation}
\label{eq:community_connectivity}
\gamma_{\strut S}  = \frac{\displaystyle \sum_{i,j \in V(S)} a_{ij}}{\displaystyle\sum_{i \in V(S),\, j \in V(G)} a_{ij}},
\end{equation}
which provides a computationally efficient and interpretable measure of subgraph cohesiveness.

The coreness values are determined iteratively. The process begins with the vertex having the smallest weighted degree, denoted as \( S_1 = \{1\} \), for which \( \gamma_1 \equiv \gamma_{S_1} = 0 \).  
For the next iteration, consider all subsets \( S_2^{(j)} = S_1 \cup \{j\} \) for \( 2 \le j \le N \), and compute \( \gamma_{S_2^{(j)}} \) for each.  
The subset \( S_2 \) corresponding to the minimum persistence probability \( \gamma_{S_2^{(k)}} \) is selected, and \( \gamma_2 \equiv \gamma_{S_2} \) is assigned as the coreness of vertex \( k \). This ensures \( \gamma_1 \le \gamma_2 \). The process continues iteratively: \( S_3 \) is formed from \( S_2 \) by adding one vertex at a time such that \( \gamma_3 \) is minimized, ensuring
\[
\gamma_1 \le \gamma_2 \le \gamma_3 \le \cdots \le \gamma_N.
\]

Thus, each vertex receives a coreness value corresponding to its inclusion in the sequence of subsets that minimizes the persistence probability. 


\subsection{Distributional comparison metrics}

To evaluate whether the CP structure differs across correlations extracted via denoising, we compare the distributions of the cp-centralization measure \(Q^{\mathrm{cp}}\) obtained from the full \(C^{\mathrm{FULL}}\), structured \(C^{\mathrm{STR}}\), and random \(C^{\mathrm{RND}}\) modes.

The two-sample Kolmogorov--Smirnov (KS) test is employed to examine whether two distributions originate from the same underlying probability law~\cite{stephens1974edf}. The KS statistic measures the maximum deviation between empirical cumulative distribution functions (ECDFs), providing a fully non-parametric measure of distributional difference without requiring normality assumptions. For two empirical distributions with ECDFs $F_1(x)$ and $F_2(x)$, the KS statistic is defined as
\begin{equation}
D = \sup_x \left| F_1(x) - F_2(x) \right|.
\end{equation}

The test evaluates the null hypothesis
\begin{equation}
H_0: F_1(x) = F_2(x), \quad \forall x,
\end{equation}
against the alternative hypothesis
\begin{equation}
H_1: F_1(x) \neq F_2(x).
\end{equation}

The interpretation of the KS statistic is intuitive: $D \approx 0$ indicates nearly identical distributions, whereas $D \approx 1$ indicates almost non-overlapping distributions. 

While the KS statistic measures the maximum difference between ECDFs, the Wasserstein distance quantifies the global distributional difference. The first Wasserstein distance~\cite{vallender1974calculation} between the distributions $F_1$ and $F_2$ is given by
\begin{equation}
W = \int_{-\infty}^{\infty} |F_1(x) - F_2(x)| dx.
\end{equation}

It can be interpreted as the minimum “transport cost” required to transform one distribution into the other. 

\subsection{Power-law characterization method}

To investigate the heterogeneity of the network topology, we analyze the tail behavior of the degree distributions. A network is said to exhibit scale-free structure if the probability distribution of vertex degrees follows a power-law form~\cite{barabasi1999emergence} 

\begin{equation}
P(k) \propto k^{-\alpha}, \quad k \ge k_{\min},
\end{equation}

where $k$ denotes the vertex degree, $\alpha$ is the scaling exponent, and $k_{\min}$ represents the lower cutoff above which the power-law behaviour holds.

The scaling exponent $\alpha$ is estimated using the maximum likelihood estimator for discrete power-law distributions,

\begin{equation}
\hat{\alpha}
=
1
+
n
\left[
\sum_{i=1}^{n}
\ln
\left(
\frac{k_i}{k_{\min}-\frac{1}{2}}
\right)
\right]^{-1},
\end{equation}

where $k_i \ge k_{\min}$ are the observed weighted degrees and $n$ is the number of samples in the tail.

The goodness-of-fit of the power-law model is evaluated using the Kolmogorov--Smirnov (KS) statistic

\begin{equation}
D
=
\max_{k \ge k_{\min}}
\left|
F_{\mathrm{emp}}(k)
-
F_{\mathrm{pl}}(k)
\right|,
\end{equation}

where $F_{\mathrm{emp}}(k)$ and $F_{\mathrm{pl}}(k)$ denote the empirical and fitted cumulative distributions, respectively. Statistical significance is assessed via a Monte Carlo procedure in which synthetic power-law samples are generated and compared with the empirical distribution. A power-law model is considered plausible when the resulting $p$-value satisfies $p>0.1$.

\subsection{Portfolio construction and performance evaluation}

We used daily log-returns of 140 stocks from the NIFTY 200, 312 stocks from the NIFTY 500 (National Stock Exchange of India), and 425 stocks from the S\&P 500 (U.S. equity market) over the period January 2010 to December 2022. This sample includes major market disruptions, such as the COVID-19 crash of 2020, ensuring a realistic and volatile market environment (see Fig.~\ref{fig:nifty}). To construct portfolios, we employ one day rolling windows of length 500 trading days. For each window, we estimate the empirical correlation matrix \(C^{\mathrm{FULL}}\) and its denoised counterpart \(C^{\mathrm{STR}}\), obtained by retaining statistically significant eigenmodes. The financial networks are then constructed from these matrices and their core--periphery (CP) structures are identified using the Markov chain–based random walk method \cite{rossa2013profiling}. The top-\(M\) core or periphery stocks are selected for portfolio construction, and out-of-sample performance is evaluated over the subsequent 125 trading days.

From a practical perspective, constructing portfolios from a large universe of assets is often associated with high transaction costs and operational complexity. Therefore, our framework focuses on selecting a relatively small subset of informative stocks. To evaluate scalability and diversification effects, we consider portfolio sizes \(M = 10, 20, 30, 40\), allowing us to examine the trade-off between diversification benefits and portfolio manageability. We consider the following portfolio strategies. Let \(\boldsymbol{\pi}^{(s)}_x\) denote a portfolio constructed from network type \(s \in \{\mathrm{FULL}, \mathrm{STR}\}\), where \(x \in \{c, p\}\) indicates whether assets are selected from the core or the periphery. Thus, \(\boldsymbol{\pi}^{(s)}_c\) and \(\boldsymbol{\pi}^{(s)}_p\) represent core- and periphery-based portfolios, respectively.

In addition to CP-based strategies, we include two benchmark portfolios. First, the \emph{highest Sharpe ratio} strategy, denoted by \(\boldsymbol{\pi}^{\mathrm{HSR}}\), selects the top-\(M\) stocks with the highest in-sample Sharpe ratios, representing a return–risk optimal stock-picking benchmark without network information. Second, the \emph{random selection} strategy, denoted by \(\boldsymbol{\pi}^{\mathrm{RND}}\), selects \(M\) stocks uniformly at random, serving as a baseline. Finally, we consider the \emph{market portfolio}, denoted by \(\boldsymbol{\pi}^{\mathrm{MKT}}\), which includes all available stocks in the respective index (140 for NIFTY 200, 312 for NIFTY 500, and 425 for S\&P 500), with weights assigned either uniformly or via Markowitz optimization. This portfolio serves as a broad benchmark for comparison.

\subsection{Monte Carlo robustness testing procedure}

To assess the robustness and statistical significance of the proposed strategy, we employ a Monte Carlo--based subsampling framework. Let \( B^{(k)} \in \mathbb{R}^{T \times W} \) denote the matrix of out-of-sample portfolio returns corresponding to strategy \( k \), where \( T = 250 \) is the investment horizon and \( W \) is the number of rolling windows. The strategies under consideration are
\[
\boldsymbol{\pi}^{FULL}_p,\;
\boldsymbol{\pi}^{STR}_p,\;
\boldsymbol{\pi}^{FULL}_c,\;
\boldsymbol{\pi}^{STR}_c,\;
\boldsymbol{\pi}^{HSR},\;
\boldsymbol{\pi}^{RND},\;
\boldsymbol{\pi}^{MKT},
\]
where \( \boldsymbol{\pi}^{STR}_p \) represents the proposed strategy.

At each iteration \( m = 1, \ldots, M \) (with \( M = 1000 \)), we randomly select a subset of \( n = 500 \) rolling windows without replacement:
\[
\mathcal{I}_m \subset \{1,\ldots,W\},
\qquad
|\mathcal{I}_m| = 500.
\]

For each strategy \( k \), the corresponding returns are pooled across the selected windows to form
\[
\mathcal{R}_m^{(k)}
=
\left\{
B^{(k)}_{t,i}
:
i \in \mathcal{I}_m,\;
t = 1,\ldots,T
\right\}.
\]

The annualized Sharpe ratio for strategy \( k \) in iteration \( m \) is computed as
\[
S_m^{(k)}
=
\frac{\mu_m^{(k)}}{\sigma_m^{(k)}}\sqrt{252},
\]
where \( \mu_m^{(k)} \) and \( \sigma_m^{(k)} \) denote the sample mean and standard deviation of \( \mathcal{R}_m^{(k)} \), respectively. The risk-free rate is assumed to be zero.

To evaluate whether the superior performance of the proposed strategy is statistically significant, we compare its Sharpe ratio with that of each benchmark across all Monte Carlo replications. Let
\begin{equation}
\Delta_m^{(j)}
=
S_m^{(\pi_p^{STR})}
-
S_m^{(j)},
\qquad
m = 1,\ldots,M,
\end{equation}
denote the Sharpe ratio difference between the proposed strategy and benchmark \( j \) in replication \( m \).

Since the distribution of Sharpe ratio differences may deviate from normality, we employ the nonparametric Wilcoxon signed-rank test. Specifically, we test
\[
H_0:\operatorname{median}\!\left(\Delta_m^{(j)}\right)=0
\]
against the one-sided alternative
\[
H_1:\operatorname{median}\!\left(\Delta_m^{(j)}\right)>0.
\]

Rejection of the null hypothesis indicates that the proposed strategy delivers significantly higher risk-adjusted performance than the corresponding benchmark.

In addition, we report the win proportion
\begin{equation}
\hat{p}^{(j)}
=
\frac{1}{M}
\sum_{m=1}^{M}
\mathbf{1}
\!\left(
S_m^{(\pi_p^{STR})}
>
S_m^{(j)}
\right),
\end{equation}
which measures the fraction of Monte Carlo replications in which the proposed strategy achieves a higher Sharpe ratio than the benchmark \( j \). Together, \( \hat{p}^{(j)} \) and the Wilcoxon p-value provide complementary evidence on the consistency and statistical significance of the observed performance.

\section{Results \& Discussion}

\subsection{Eigenvalue spectrum and structured-mode extraction}

\begin{figure}[htbp]
    \centering

    \begin{subfigure}[b]{0.39\textwidth}
        \centering
        \includegraphics[width=\textwidth]{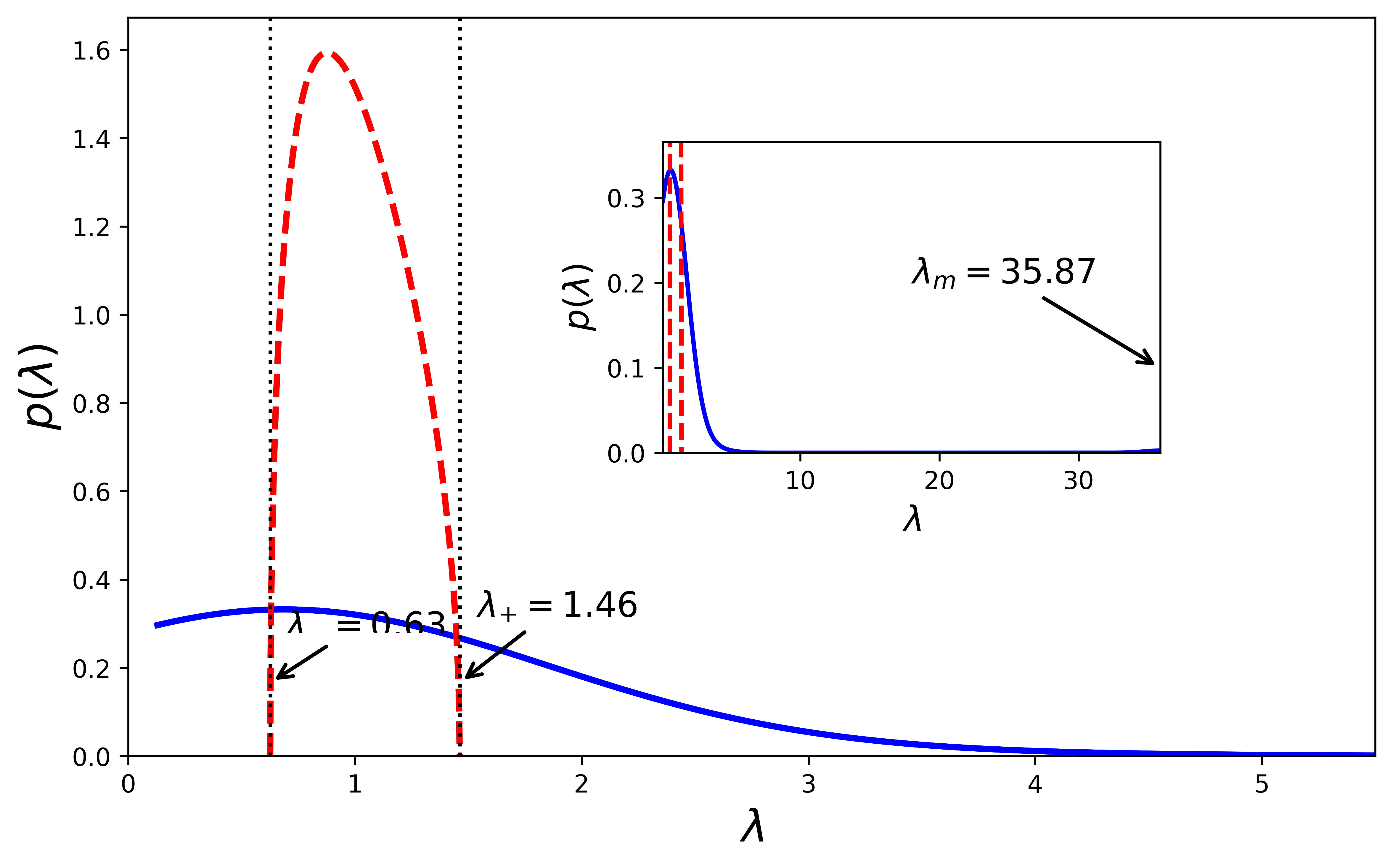}
        \caption{}
        \label{fig:nifty200}
    \end{subfigure}
    \hfill
    \begin{subfigure}[b]{0.39\textwidth}
        \centering
        \includegraphics[width=\textwidth]{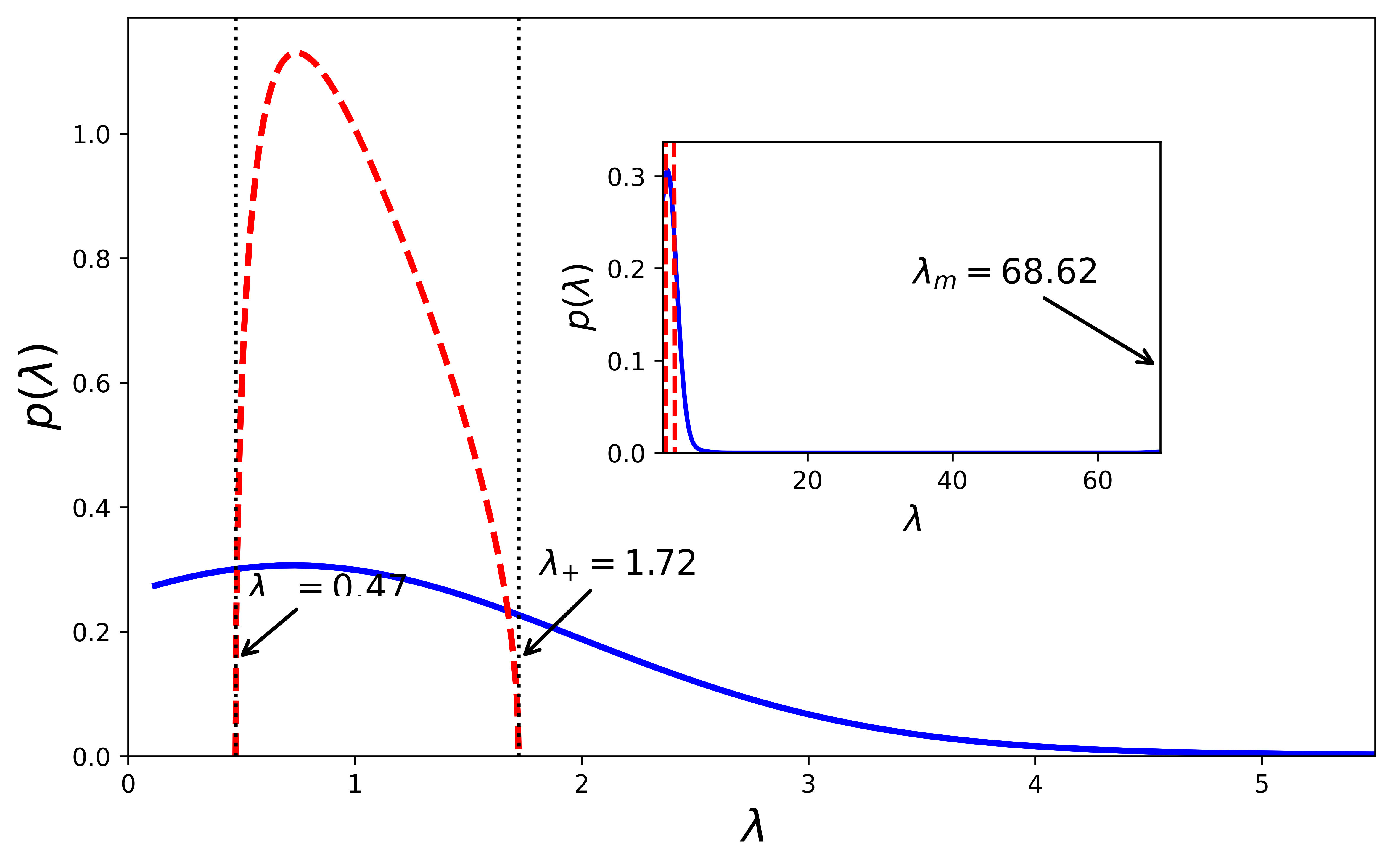}
        \caption{}
        \label{fig:nifty500}
    \end{subfigure}

    \vspace{0.6cm} 

    \begin{subfigure}[b]{0.39\textwidth}
        \centering
        \includegraphics[width=\textwidth]{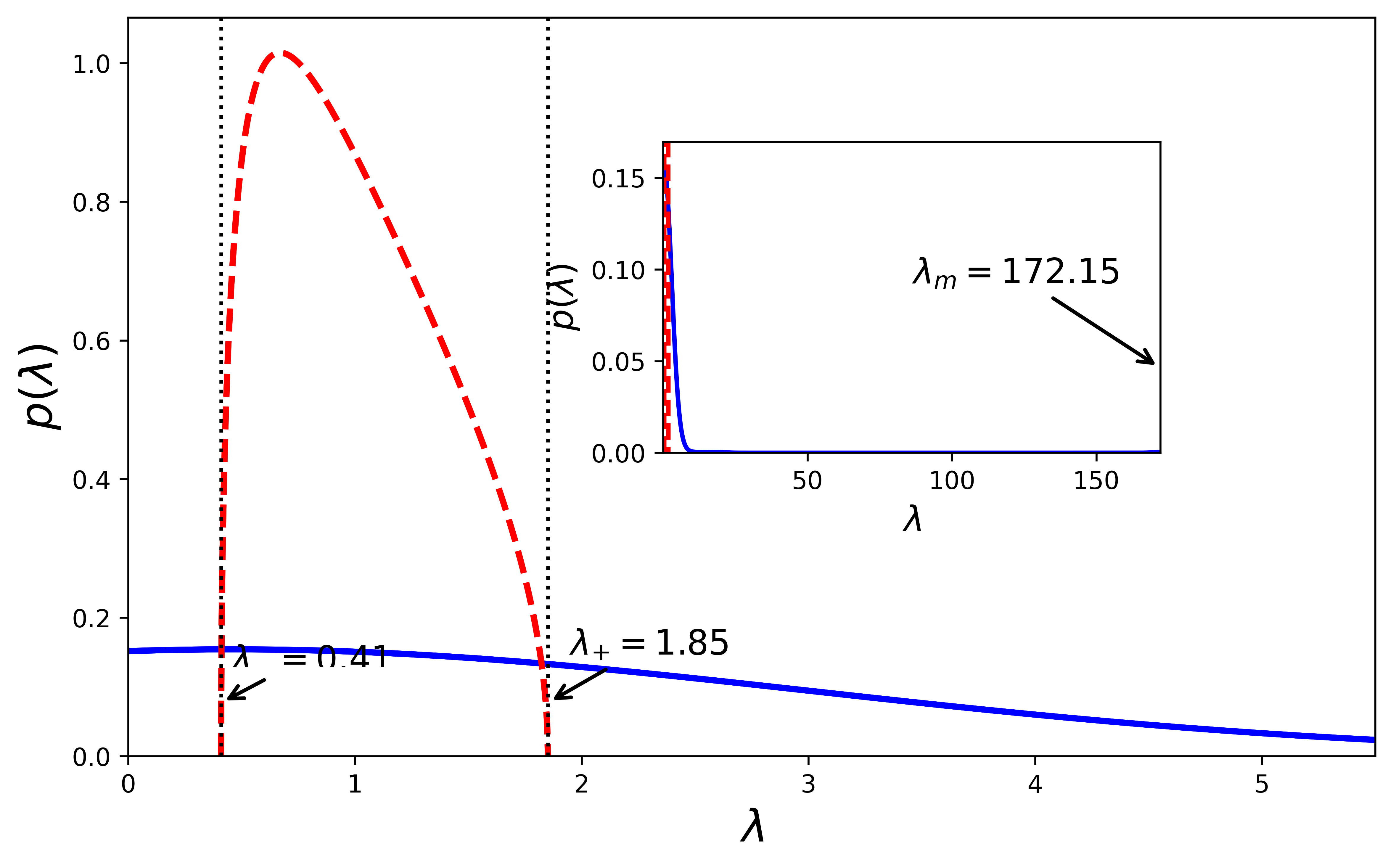} 
        \caption{}
        \label{fig:sp500}
    \end{subfigure}

\caption{Eigenmode plots for (a) NIFTY~200, (b) NIFTY~500, and (c) S\&P~500 over the period from January~1,~2010, to December~31,~2022. Here, $\lambda_{-}$ and $\lambda_{+}$ denote the Marčenko--Pastur (MP) bounds, and $\lambda_{m}$ represents the largest eigenvalue of the empirical correlation matrix.}

    \label{fig:eigen_modes_plot}
\end{figure}

\begin{figure}[htbp]
    \centering
    \begin{subfigure}[b]{0.42\textwidth}
        \centering
        \includegraphics[width=\textwidth]{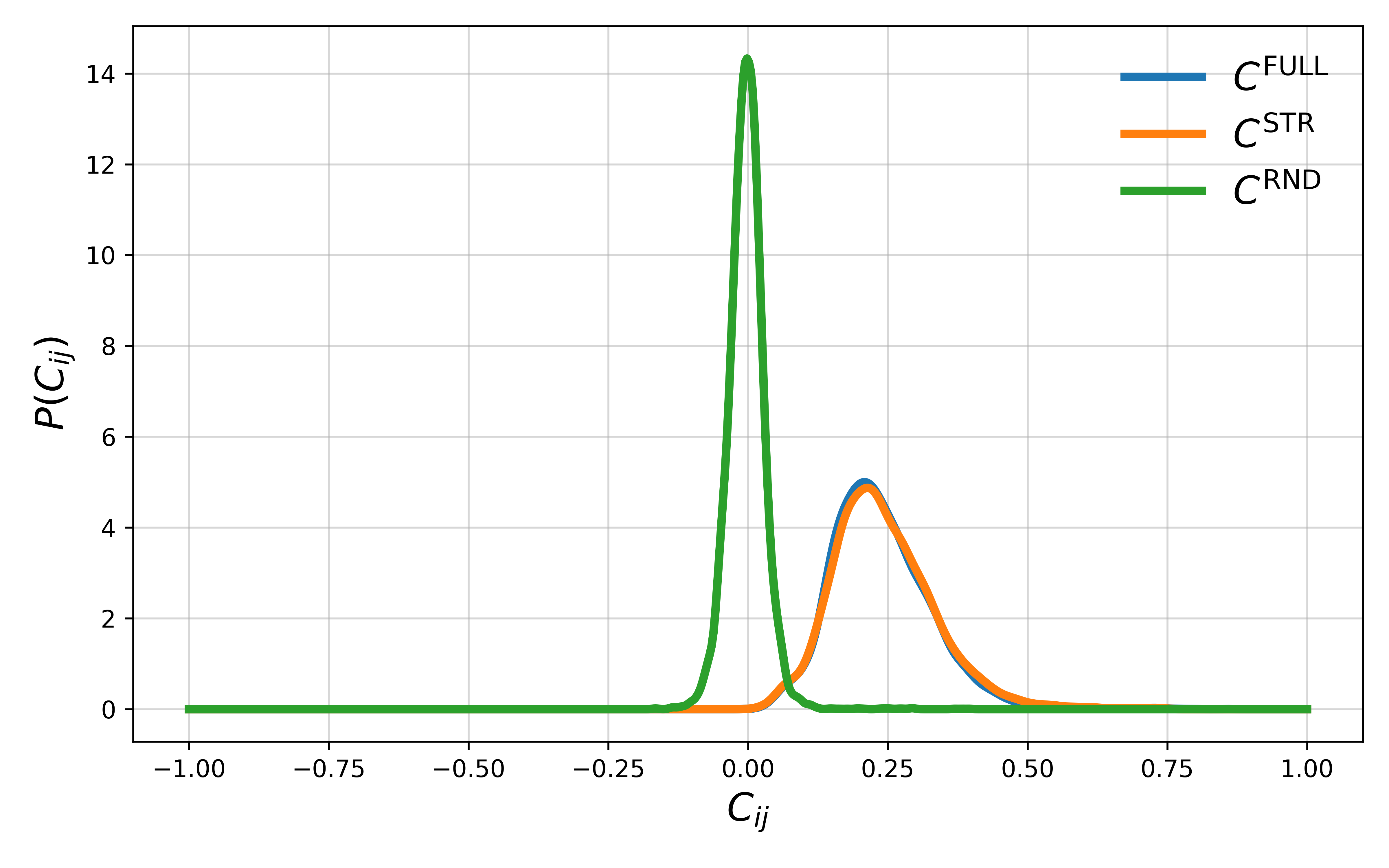}
        \caption{}
        \label{fig:nifty200}
    \end{subfigure}
    \hfill
    \begin{subfigure}[b]{0.42\textwidth}
        \centering
        \includegraphics[width=\textwidth]{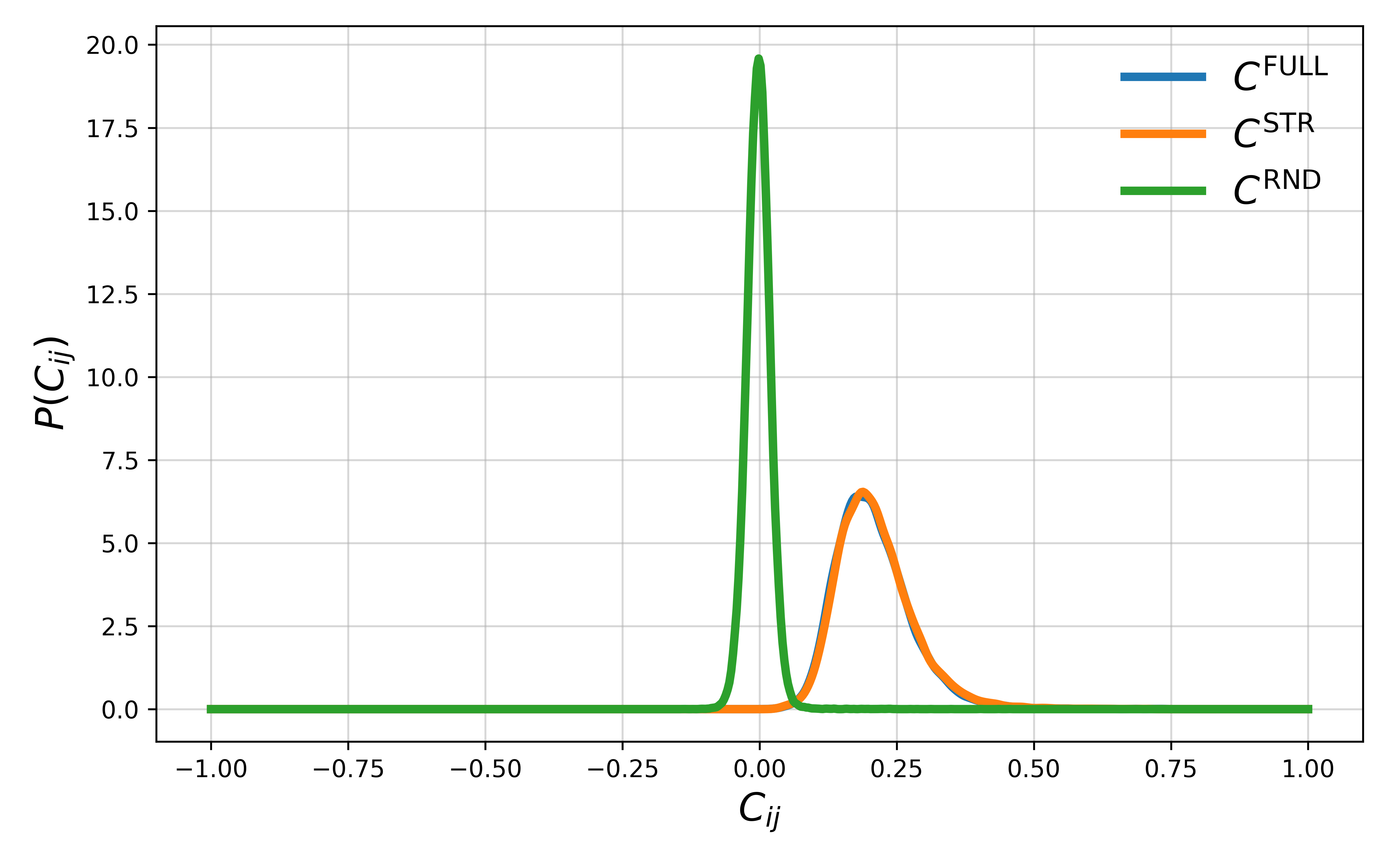}
        \caption{}
        \label{fig:nifty500}
    \end{subfigure}

    \vspace{0.6cm} 

    \begin{subfigure}[b]{0.42\textwidth}
        \centering
        \includegraphics[width=\textwidth]{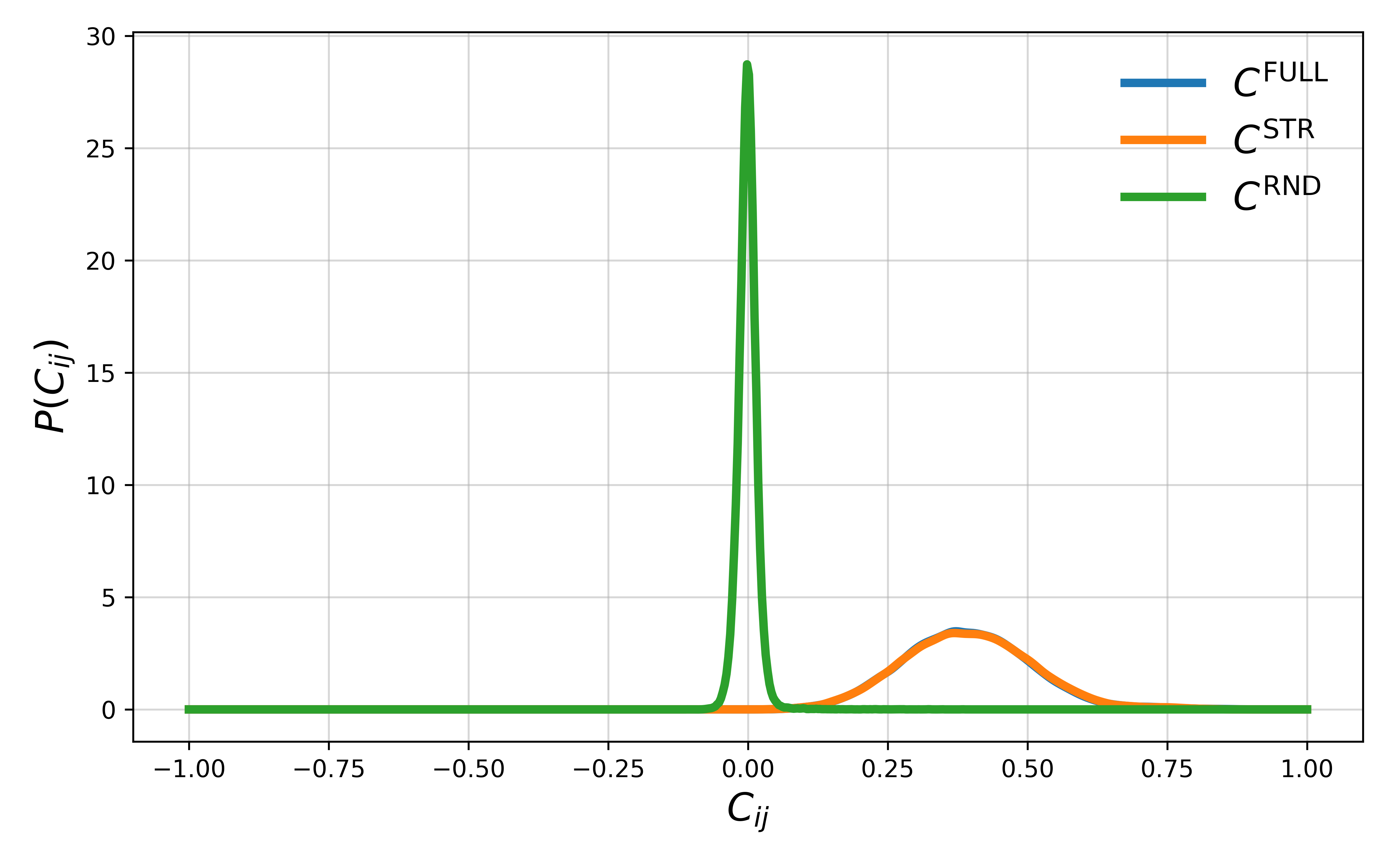} 
        \caption{}
        \label{fig:sp500}
    \end{subfigure}

\caption{Kernel density estimates of the empirical distribution of pairwise correlation coefficients ($C_{ij}$) for (a) NIFTY 200, (b) NIFTY 500, and (c) S\&P 500 over the period January 1, 2010 to December 31, 2022.}    
    \label{fig:corr_distribution}
\end{figure}

Throughout the 13-year dataset, we find that for the NIFTY~200 index, the 10 largest eigenvalues fall within the structured mode, while for the NIFTY~500 and S\&P~500 indices, 12 and 16 eigenvalues are identified, respectively. Table~\ref{tab:rmt_spectral} shows the spectral properties of empirical correlation matrix. Figure~\ref{fig:eigen_modes_plot} presents the eigenmode plots for the NIFTY~200, NIFTY~500, and S\&P~500. Subsequently, networks are constructed from the full cross-correlation matrix $C^{\mathrm{FULL}}$, the structured-mode matrix $C^{\mathrm{STR}}$, and the random-mode matrix $C^{\mathrm{RND}}$. Figure~\ref{fig:corr_distribution} shows the probability density functions of the elements of the correlation matrices for the NIFTY~200, NIFTY~500, and S\&P~500. The distributions corresponding to the full and structured-mode correlation matrices are in close agreement, highlighting that the structured-mode matrix constructed using only a few eigenvalues lying beyond the Marchenko--Pastur bounds captures the dominant correlation structure of the market.

\begin{table}[htbp]
\centering
\caption{Spectral properties of empirical correlation matrix. 
$\lambda_{-}$ and $\lambda_{+}$ denote the theoretical Marčenko--Pastur (MP) bounds and $\lambda_{\max}$ is the largest eigenvalue of the empirical correlation matrix. 
Percentages indicate the proportion of empirical eigenvalues ($\hat{\lambda}$) lying below $\lambda_{-}$, within the MP bulk $(\lambda_{-}<\hat{\lambda}<\lambda_{+})$, and above $\lambda_{+}$.}
\label{tab:rmt_spectral}
\resizebox{\textwidth}{!}{
\begin{tabular}{l c c c c c c}
\hline
Dataset & $\lambda_{-}$ & $\lambda_{+}$ & $\lambda_{\max}$ & 
$\hat{\lambda}<\lambda_{-}$ (\%) & 
$\lambda_{-}<\hat{\lambda}<\lambda_{+}$ (\%) & 
$\hat{\lambda}>\lambda_{+}$ (\%) \\
\hline

NIFTY 200 & 0.626 & 1.461 & 35.87 & 46.43 & 46.43 & 7.14 \\

NIFTY 500 & 0.473 & 1.722 & 68.62 & 25.00 & 71.15 & 3.85 \\

S\&P 500 & 0.409 & 1.851 & 172.15 & 57.65 & 38.59 & 3.76 \\

\hline
\end{tabular}
}
\end{table}

\begin{figure}[htbp]
    \centering

    \begin{subfigure}[b]{1.0\textwidth}
        \centering
        \includegraphics[width=1.0\textwidth]{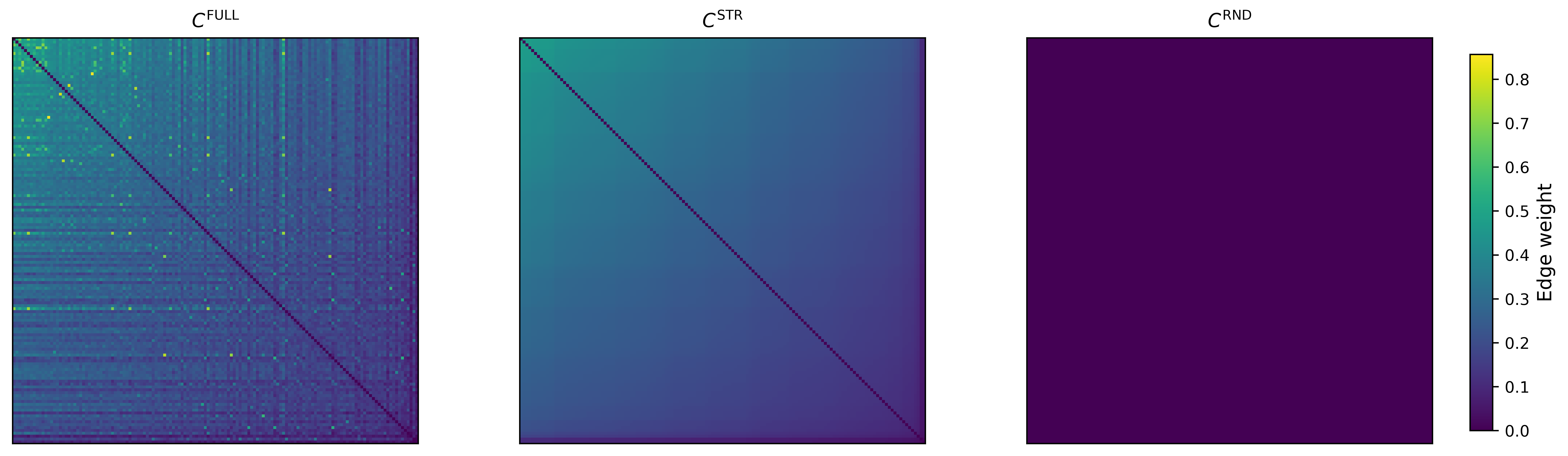}
        \caption{NIFTY 200}
        \label{fig:adjacency_nifty200}
    \end{subfigure}
    
    \vspace{0.3cm} 
    
    \begin{subfigure}[b]{1.0\textwidth}
        \centering
        \includegraphics[width=1.0\textwidth]{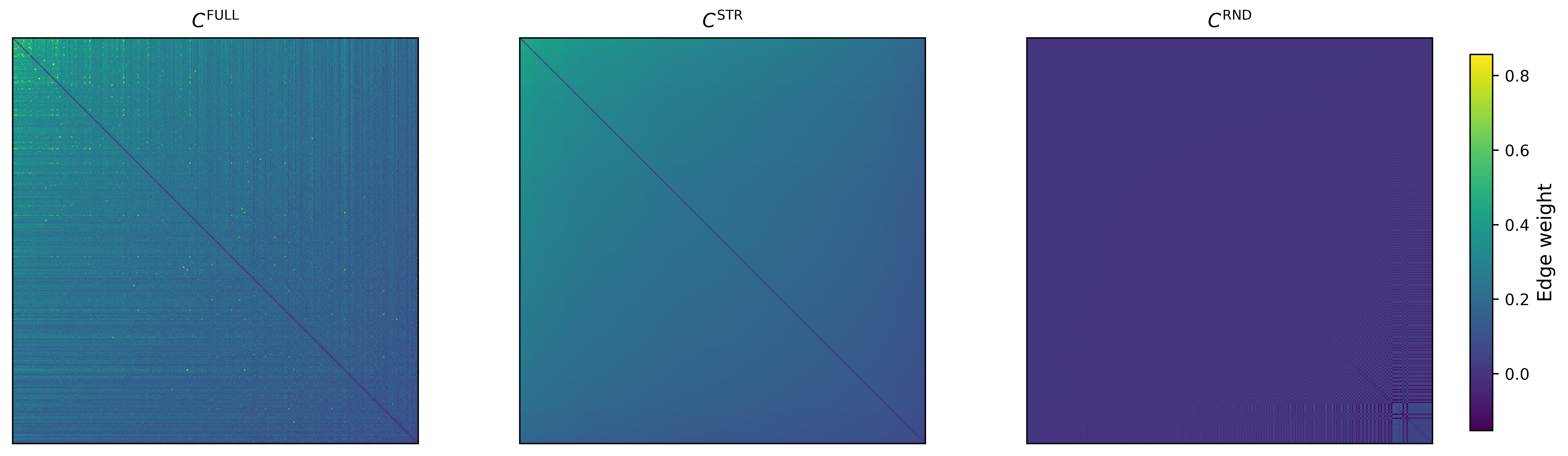}
        \caption{NIFTY 500}
        \label{fig:adjacency_nifty500}
    \end{subfigure}

    \vspace{0.3cm} 
    
    \begin{subfigure}[b]{1.0\textwidth}
        \centering
        \includegraphics[width=1.0\textwidth]{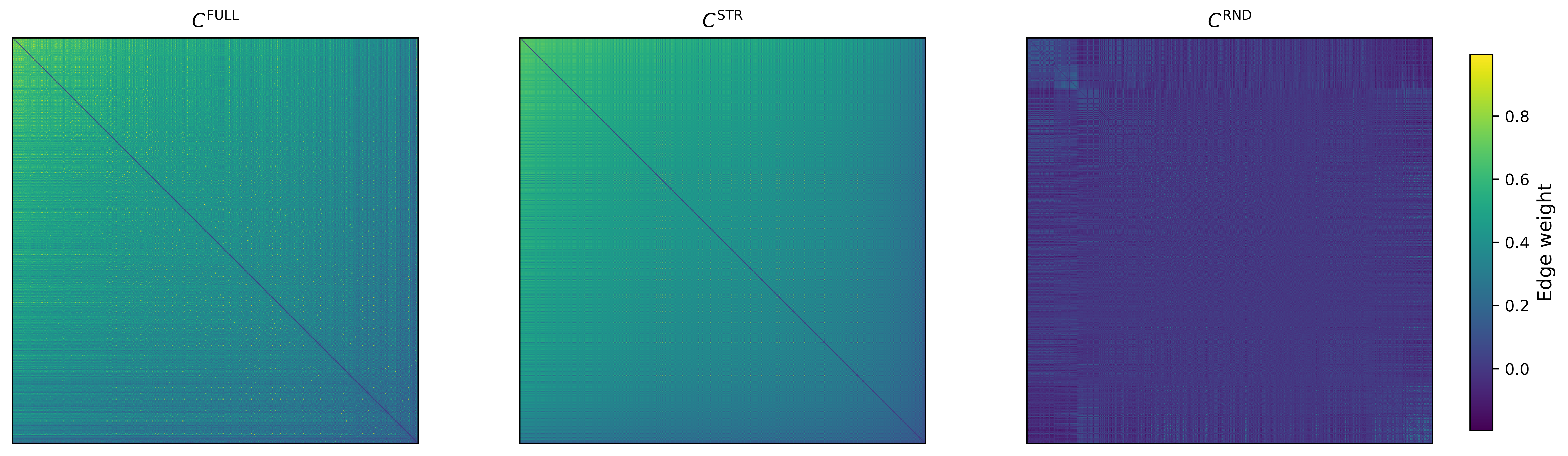}
        \caption{S\&P 500}
        \label{fig:adjacency_sp500}
    \end{subfigure}

    \caption{Permuted adjacency matrix plots of NIFTY 200, NIFTY 500, and S\&P 500 indices.}
    \label{fig:adjacency_plot}
\end{figure}

In Figure~\ref{fig:adjacency_plot}, we plot the permuted adjacency matrices from the full ($C^{\mathrm{FULL}}$), structured ($C^{\mathrm{STR}}$), and random ($C^{\mathrm{RND}}$) mode matrices for the NIFTY 200, NIFTY 500, and S\&P 500 over the entire 13-year period. The matrices are permuted by ordering rows and columns in decreasing order of coreness values. A clear core--periphery block structure is observed for the full and structured-mode networks; however, no such structure is evident for the random-mode network.

\subsection{Core–periphery organization: full vs. structured vs. random networks}

In this section, we examine the stability of the detected CP structure using the full, structured, and random mode correlation matrices for the NIFTY 200, NIFTY 500, and S\&P 500 indices. The networks are constructed using a rolling window of one day for each correlation mode. For each network, we construct the minimum spanning tree using the standard Euclidean distance derived from the full and mode cross-correlation matrices to filter complex datasets by extracting a network of representative links~\cite{mantegna1999hierarchical}.

\begin{figure}[htbp]
    \centering

    \begin{subfigure}[b]{0.48\textwidth}
        \centering
        \includegraphics[width=\textwidth]{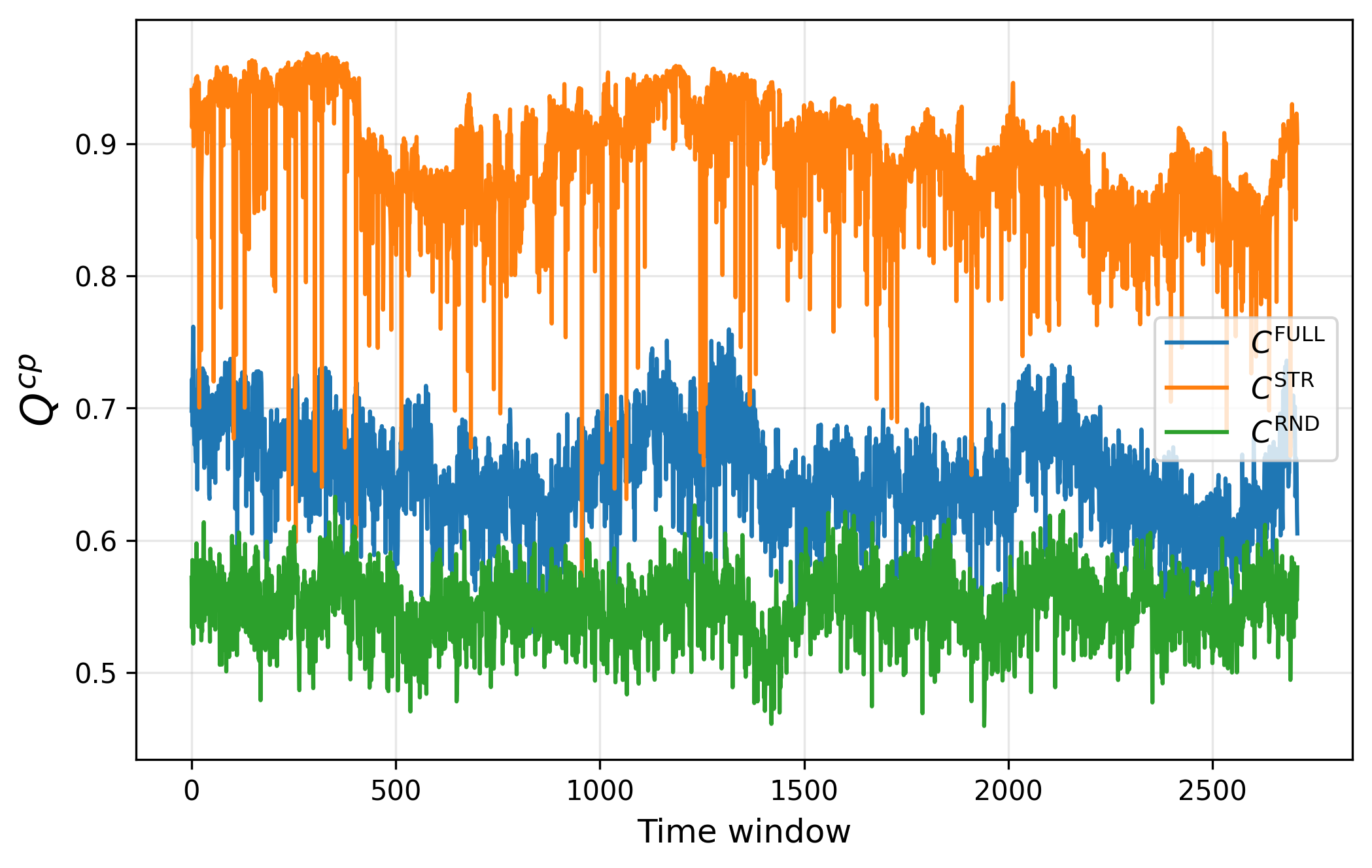}
        \caption{}
        \label{fig:nifty200}
    \end{subfigure}
    \hfill
    \begin{subfigure}[b]{0.48\textwidth}
        \centering
        \includegraphics[width=\textwidth]{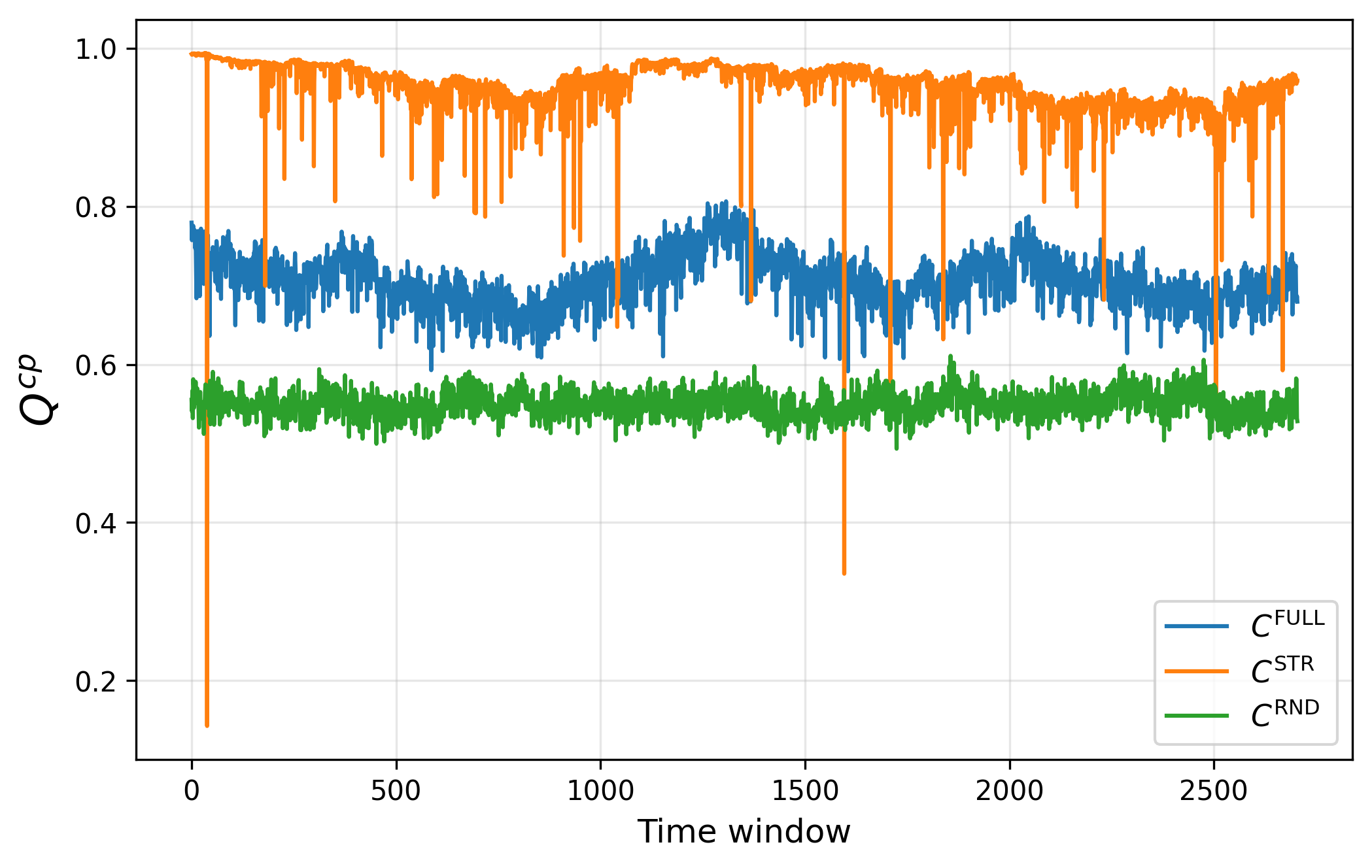}
        \caption{}
        \label{fig:nifty500}
    \end{subfigure}

    \vspace{0.6cm} 

    \begin{subfigure}[b]{0.48\textwidth}
        \centering
        \includegraphics[width=\textwidth]{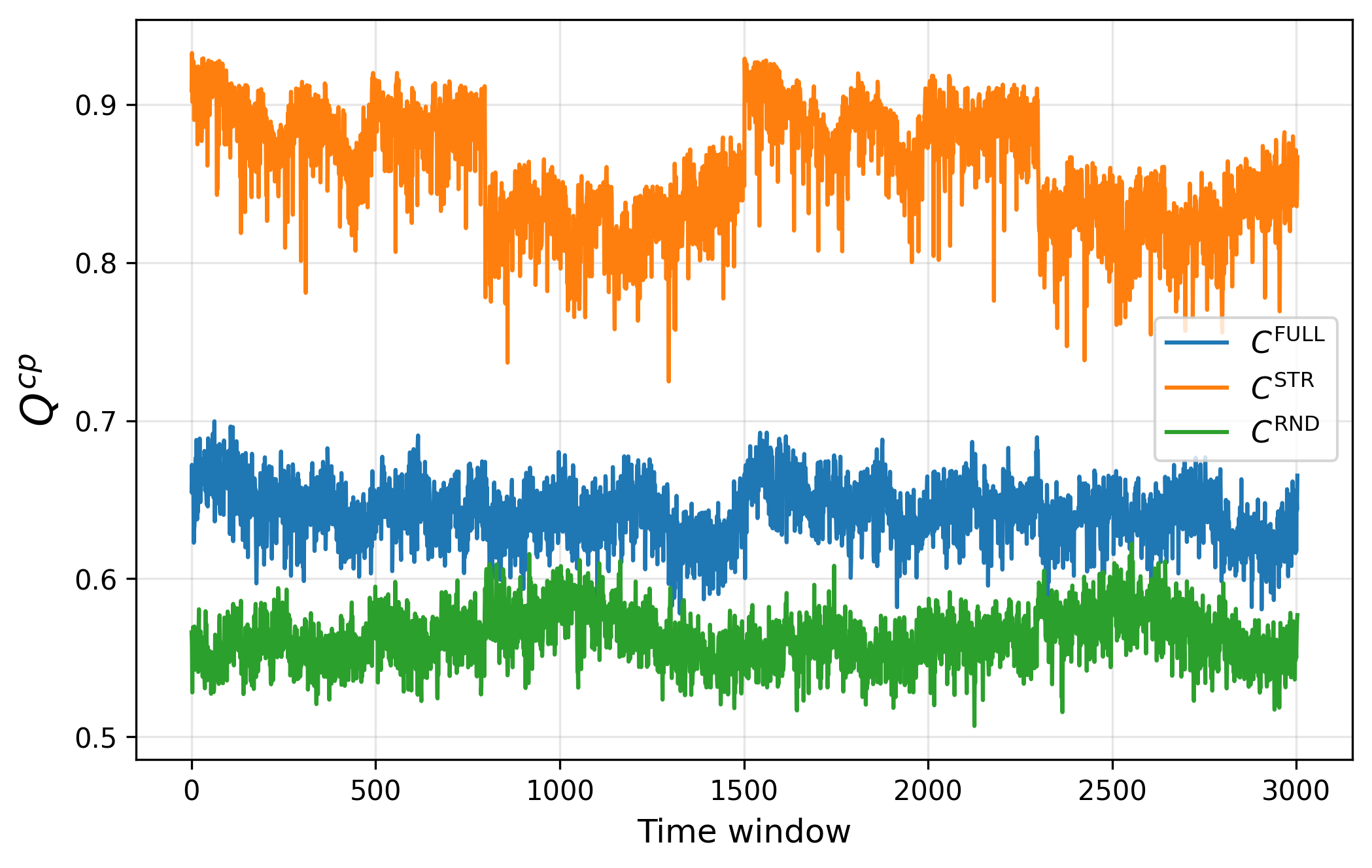} 
        \caption{}
        \label{fig:sp500}
    \end{subfigure}

\caption{$Q^{cp}$ plots for (a) NIFTY~200, (b) NIFTY~500, and (c) S\&P~500 over rolling windows.}
    \label{fig:Qcp_timeseries}
\end{figure}

For every window, we compute the core-periphery centralization index ($Q^{\mathrm{cp}}$)~\cite{rossa2013profiling}, defined as
\begin{equation}
Q^{\mathrm{cp}} = 1 - \frac{2}{n-2} \sum_{k=1}^{n-1} \gamma_k,
\end{equation}
where $(\gamma_1, \gamma_2, \ldots, \gamma_{n-1})$ denotes the CP profile of the network. Higher values of $Q^{\mathrm{cp}}$ indicate a more pronounced CP organization, whereas lower values correspond to networks with weak or ill-defined CP structures.

 Figures~\ref{fig:Qcp_timeseries} and \ref{fig:Qcp_histogram} show the temporal evolution of $Q^{\mathrm{cp}}$ for the indices of NIFTY 200, NIFTY 500 and S\&P 500. For the NIFTY 200, $Q^{\mathrm{cp}}$ exhibits values in the range $0.52$--$0.76$ for the full mode, $0.55$--$0.96$ for the structured mode and $0.45$--$0.63$ for the random mode. In the case of the NIFTY 500, the corresponding ranges are $0.59$--$0.80$, $0.14$--$0.99$, and $0.49$--$0.61$, respectively. For the S\&P 500 index, $Q^{\mathrm{cp}}$ spans $0.57$--$0.69$ for the full mode, $0.72$--$0.93$ for the structured mode, and $0.50$--$0.62$ for the random mode. Notably, the structured mode consistently exhibits a wider range of CP centralization compared to the full and random correlation networks, highlighting the stronger and more heterogeneous core-periphery organization captured by the structured mode.

\subsection{Statistical significance of core–periphery structure}

To evaluate the statistical significance of the observed CP structure, we adopt a Monte Carlo simulation framework based on degree-preserving randomization. For each empirical network, we generate 100 randomized counterparts using the configuration model~\cite{newman2003structure}, thereby preserving the original degree sequence. The cp-centralization \(Q^{cp}\) is computed for each randomized network, denoted by \(Q^{cp}_{i,\mathrm{rand}}\), \(i=1,\ldots,100\). The statistical significance is assessed using an empirical \(p\)-value defined as:

\begin{equation}
p = \frac{1}{100}\sum_{i=1}^{100} 
\mathbb{I}\!\left(Q^{cp}_{i,\mathrm{rand}} > Q^{cp}\right),
\end{equation}

which measures the probability of observing a cp-centralization larger than the empirical value under random conditions. The following hypotheses are tested:

 \textit{Null hypothesis} (\(H_0\)): The observed cp-centralization \(Q^{cp}\) arises from random network structure.
 
\textit{Alternative hypothesis} (\(H_a\)): The observed cp-centralization \(Q^{cp}\) is statistically significant and not attributable to random chance.

A sufficiently small $p$-value (e.g., $p < 0.01$) leads to the rejection of the null hypothesis $H_0$, indicating that the detected CP structure is statistically significant rather than arising from random fluctuations. For the NIFTY 200 index, the proportion of statistically significant windows at the significance level $1\%$ is $100.00\%$ for full correlation networks, $98.60\%$ for the structured mode, and $100.00\%$ for the random mode. For the NIFTY 500 index, the fraction of statistically significant windows at the level $1\%$ is $100.00\%$ for full correlation networks, $98.56\%$ for the structured mode and $100.00\%$ for the random mode. For the S\&P 500 index, the full mode is statistically significant in $100.00\%$ of the windows at the $1\%$ significance level, followed by the structured mode ($97.63\%$) and the random mode ($100.00\%$). Similar conclusions are obtained when using a $5\%$ significance level. In general, these results provide strong evidence that the observed CP structures in all analyzed datasets and network construction methods are statistically significant.

\subsection{Distributional separation between structured and random modes}

The empirical KS statistics obtained for the NIFTY 200, NIFTY 500 and S\&P 500 rolling-window networks are extremely large and are presented in Table~\ref{tab:ks_wasserstein}. These values indicate near-complete distributional separation, particularly between structured and random modes. The effectively zero p-values arise because the sample size is large and the KS statistics are extremely high, causing asymptotic p-values to numerically underflow to zero. This behavior is expected and indicates overwhelming statistical evidence against the null hypothesis.

The computed Wasserstein distances are presented in Table~\ref{tab:ks_wasserstein}. The Wasserstein distance analysis further confirms the strong distributional separation between structured and random eigenmodes. The largest distance is observed between structured and random modes, reinforcing the interpretation that structured eigenmodes capture genuine market organization, whereas random bulk eigenmodes primarily represent noise. The moderate distance between the full and structured modes indicates that the structured mode explains a substantial fraction of the full network topology. The distributional separation is further supported by summary statistics which are presented in Table~\ref{tab:summary_stats_dataset_col}

\begin{table}[htbp]
\scriptsize
\centering
\caption{Comparison of $Q^{cp}$ distributions over the rolling windows. For each pair, the Kolmogorov--Smirnov statistic $D$ (with p-value in parentheses) and the Wasserstein distance $W$ are reported.}
\label{tab:ks_wasserstein}
\begin{tabular}{llccc}
\hline
Comparison & Metric & NIFTY 200 & NIFTY 500 & S\&P 500 \\
\hline

\multirow{2}{*}{Full vs Structured}
 & KS Statistic $D$ & 0.982 ($p < 10^{-16}$) & 0.991 ($p < 10^{-16}$) & 1.000 ($p < 10^{-16}$) \\
 & Wasserstein $W$ & 0.237 & 0.245 & 0.217 \\

\hline
\multirow{2}{*}{Full vs Random}
 & KS Statistic $D$ & 0.893 ($p < 10^{-16}$) & 0.999 ($p < 10^{-16}$) & 0.970 ($p < 10^{-16}$) \\
 & Wasserstein $W$ & 0.100 & 0.157 & 0.079 \\

\hline
\multirow{2}{*}{Structured vs Random}
 & KS Statistic $D$ & 0.998 ($p < 10^{-16}$) & 0.998 ($p < 10^{-16}$) & 1.000 ($p < 10^{-16}$) \\
 & Wasserstein $W$ & 0.337 & 0.402 & 0.297 \\

\hline
\end{tabular}
\end{table}

\begin{table}[htbp]
\scriptsize
\centering
\caption{Summary statistics of $Q^{cp}$ over the rolling windows.}
\label{tab:summary_stats_dataset_col}
\begin{tabular}{llccc}
\hline
Dataset & Method & Mean & Median & Std Dev \\
\hline

\multirow{3}{*}{NIFTY 200}
 & $C^{\mathrm{FULL}}$ & 0.651 & 0.651 & 0.038 \\
 & $C^{\mathrm{STR}}$  & \textbf{0.888} & \textbf{0.894} & 0.051 \\
 & $C^{\mathrm{RND}}$  & 0.551 & 0.551 & 0.026 \\

\hline
\multirow{3}{*}{NIFTY 500}
 & $C^{\mathrm{FULL}}$ & 0.706 & 0.705 & 0.033 \\
 & $C^{\mathrm{STR}}$  & \textbf{0.951} & \textbf{0.958} & 0.040 \\
 & $C^{\mathrm{RND}}$  & 0.549 & 0.549 & 0.016 \\

\hline
\multirow{3}{*}{S\&P 500}
 & $C^{\mathrm{FULL}}$ & 0.643 & 0.643 & 0.019 \\
 & $C^{\mathrm{STR}}$  & \textbf{0.860} & \textbf{0.860} & 0.036 \\
 & $C^{\mathrm{RND}}$  & 0.563 & 0.562 & 0.017 \\

\hline
\end{tabular}
\end{table}

\begin{figure}[htbp]
    \centering

    \begin{subfigure}[b]{0.48\textwidth}
        \centering
        \includegraphics[width=\textwidth]{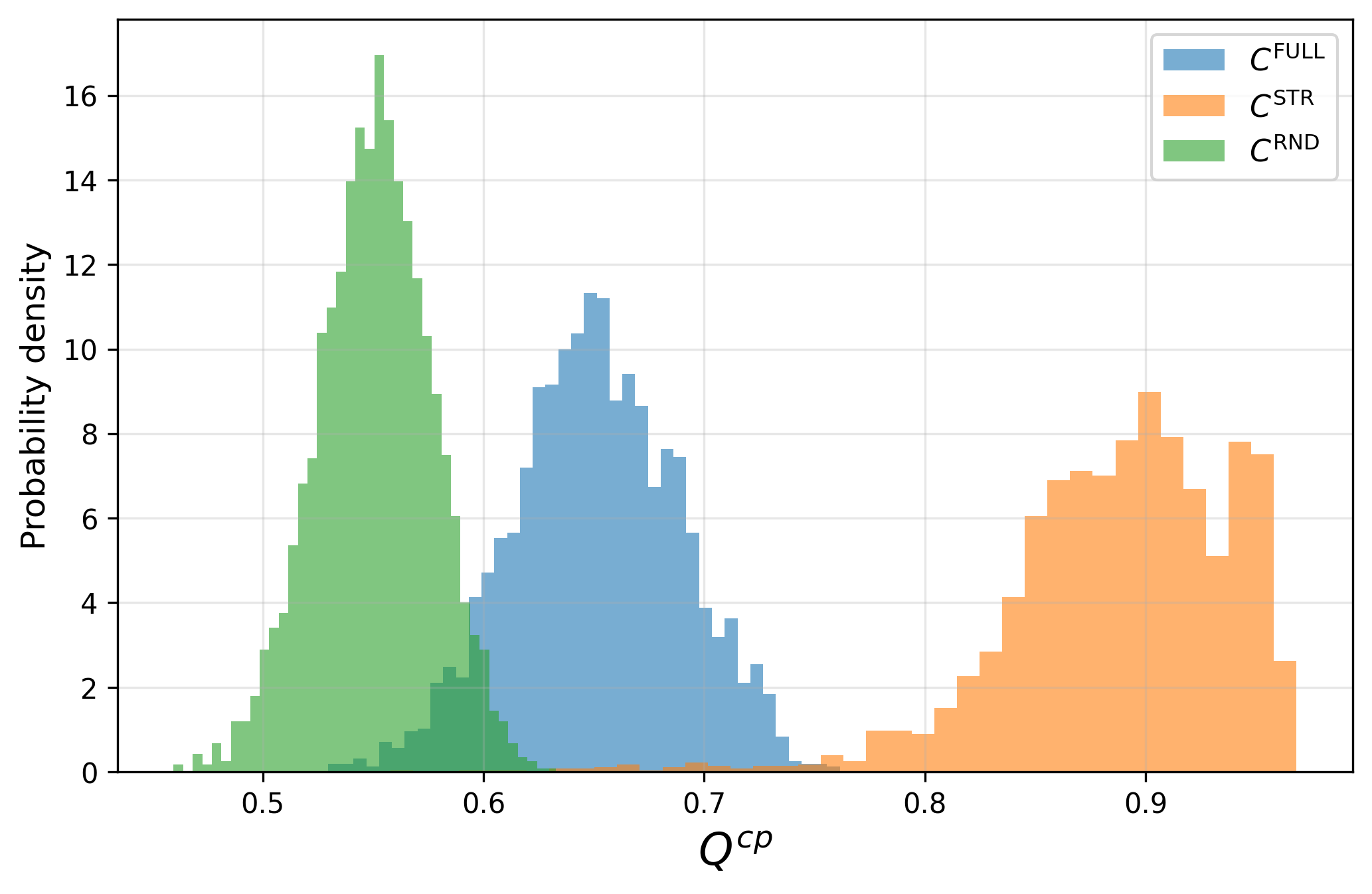}
        \caption{}
        \label{fig:nifty200}
    \end{subfigure}
    \hfill
    \begin{subfigure}[b]{0.48\textwidth}
        \centering
        \includegraphics[width=\textwidth]{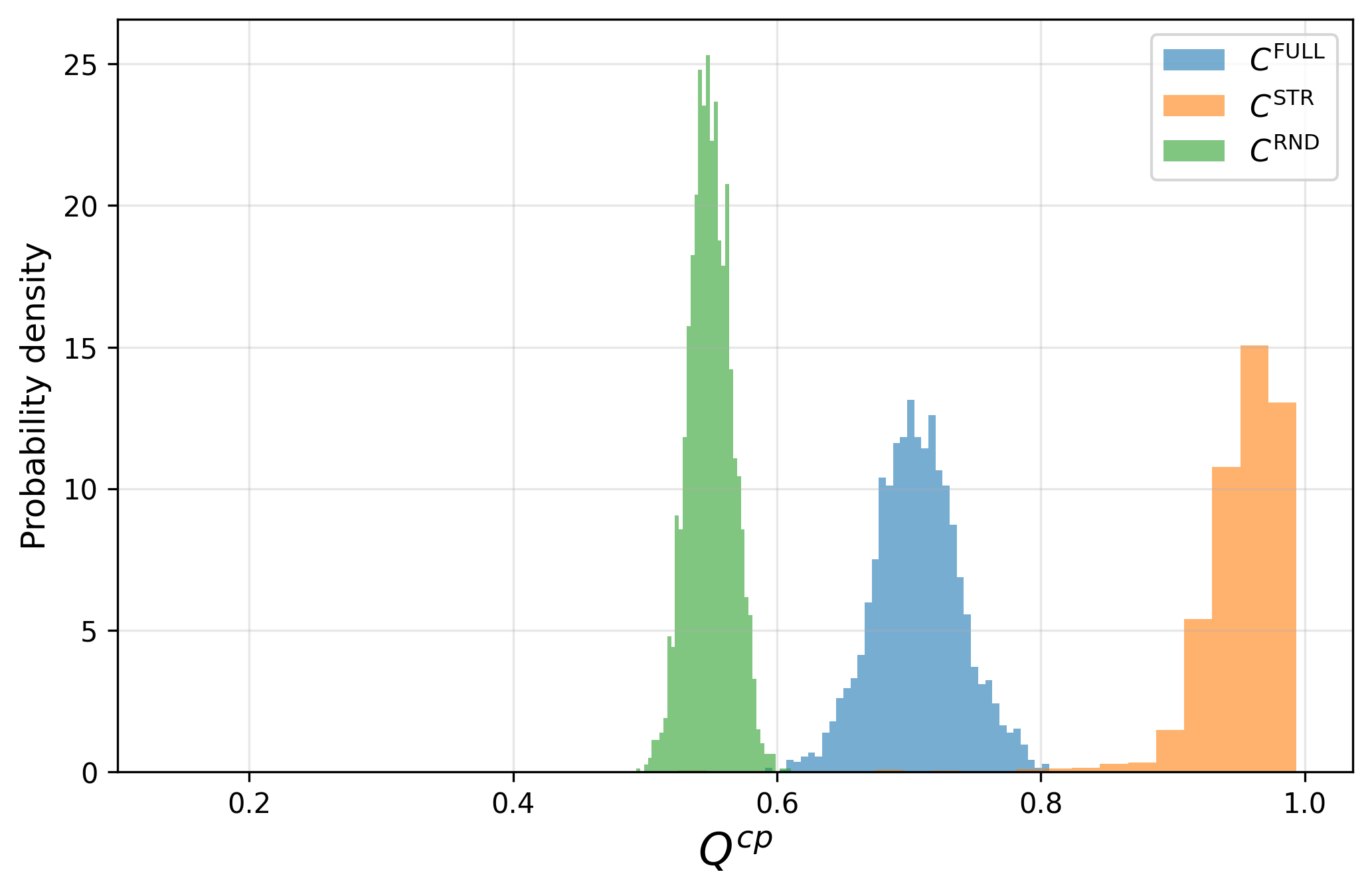}
        \caption{}
        \label{fig:nifty500}
    \end{subfigure}

    \vspace{0.6cm} 

    \begin{subfigure}[b]{0.48\textwidth}
        \centering
        \includegraphics[width=\textwidth]{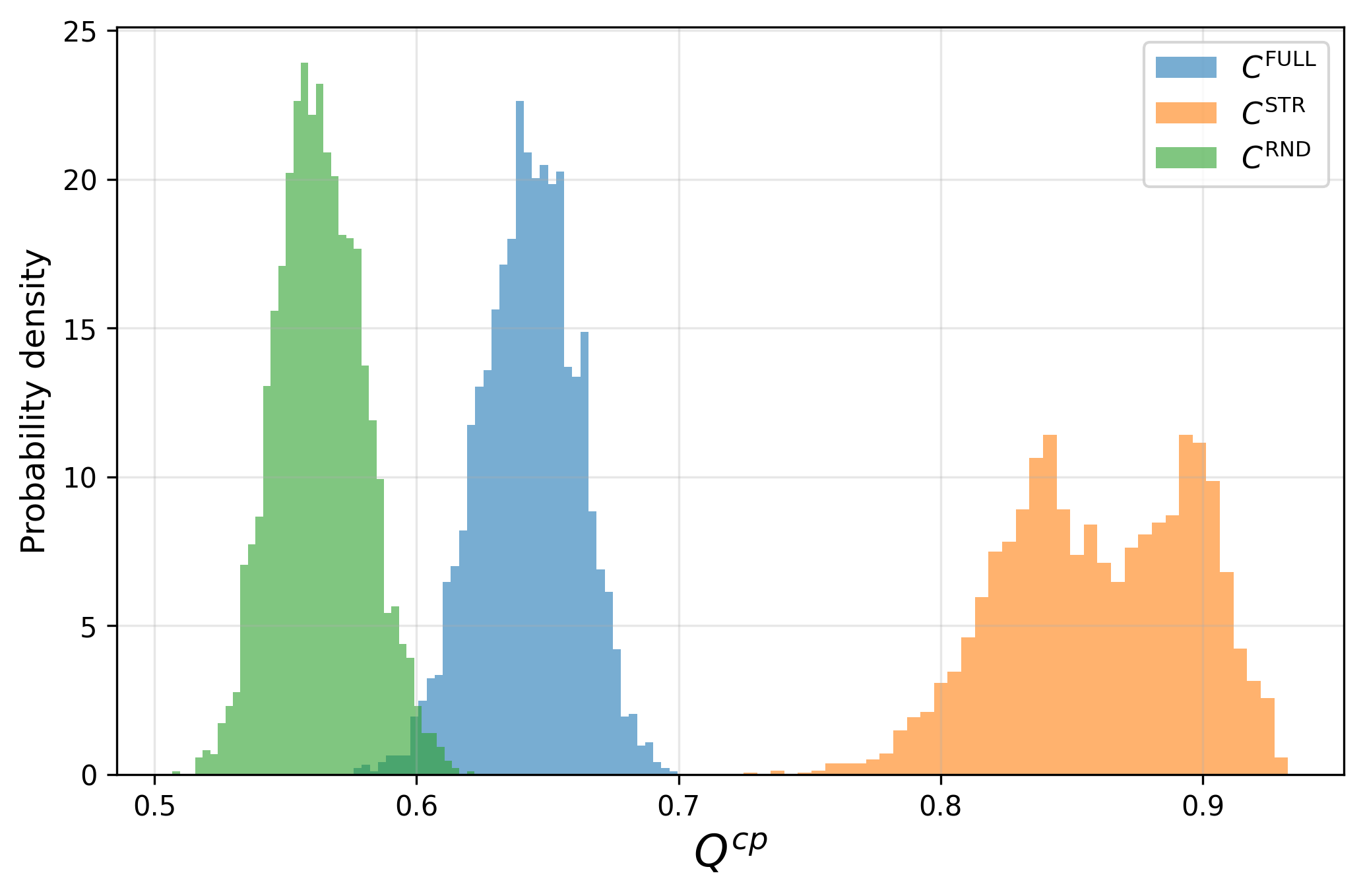} 
        \caption{}
        \label{fig:sp500}
    \end{subfigure}

\caption{$Q^{cp}$ histogram for (a) NIFTY~200, (b) NIFTY~500, and (c) S\&P~500 over the rolling windows.}
    \label{fig:Qcp_histogram}
\end{figure}

Figure~\ref{fig:Qcp_histogram} show histograms and ECDF analysis reveal that the distributions are almost non-overlapping, especially between $C^{\mathrm{STR}}$ and $C^{\mathrm{RND}}$. From a financial network perspective, this result provides strong evidence that the CP structure is primarily driven by information-bearing eigenmodes rather than random correlations. The structured mode ($C^{\mathrm{STR}}$), which captures the effects of market-wide and sectoral synchronization, produces a significantly stronger hierarchical organization. The full cross--correlation matrix ($C^{\mathrm{FULL}}$) reflects a mixture of structured information and noise, while the random mode ($C^{\mathrm{RND}}$) produces weaker and less stable CP organization consistent with noise-dominated correlations. The near-perfect separation between structured and random modes ($D \approx 1$) indicates that genuine economic structure and random noise generate fundamentally different topologies of financial networks. In conclusion, these results strongly support the hypothesis that structured eigenmodes encode genuine market information, whereas random bulk eigenmodes primarily represent noise.

\subsection{Scale-free characterization of denoised networks}

The estimated scaling exponents and goodness-of-fit statistics for the vertex strength distributions are summarized in Table~\ref{tab:powerlaw_compare}. The results reveal clear differences between the full, structured, and random mode networks in the three markets. For the NIFTY 200 dataset, only the structured network exhibits statistically consistent power-law behaviour ($p>0.1$), with a scaling exponent in the range $2<\alpha<3$, indicating a heterogeneous hub-dominated organization. In contrast, both the full and random mode networks significantly deviate from a power-law model. For the NIFTY 500 dataset, both the full and structured networks are consistent with power-law behaviour, with the structured network showing the strongest agreement with the scaling model. The random network again strongly rejects the power-law hypothesis. In contrast, none of the networks for the S\&P 500 dataset exhibit statistically significant power-law behaviour, indicating that the U.S. market is less consistent with a scale-free organization. These results indicate that eigenmode filtering reveals a clearer scale-free structure of the Indian markets, whereas the random eigenmodes do not exhibit scale-free organization.

\begin{table}[H]
\scriptsize
\centering
\caption{Power-law fitting results for vertex strength distributions of MST networks. For each mode, the scaling exponent $\alpha$ (with $p$-value in parentheses) 
and the lower cutoff $k_{\min}$ are reported.}
\label{tab:powerlaw_compare}

\begin{tabular}{llccc}

\hline
Mode & Metric & NIFTY 200 & NIFTY 500 & S\&P 500 \\
\hline

\multirow{2}{*}{Full}
& $\alpha$ ($p$-value)
& 1.94 $(6.83\times10^{-13})$
& 2.38 $(0.146)$
& 2.52 $(1.70\times10^{-4})$

\\

& $k_{\min}$
& 1.165
& 3.534
& 2.667

\\
\hline

\multirow{2}{*}{Structured}
& $\alpha$ ($p$-value)
& \textbf{2.78} $(0.452)$
& \textbf{2.60} $(0.526)$
& 1.71 $(3.14\times10^{-4})$

\\

& $k_{\min}$
& 5.790
& 6.768
& 1.083

\\
\hline

\multirow{2}{*}{Random}
& $\alpha$ ($p$-value)
& 1.86 $(1.22\times10^{-30})$
& 1.85 $(3.68\times10^{-67})$
& 1.92 $(4.14\times10^{-68})$

\\

& $k_{\min}$
& 5.790
& 6.768
& 1.083

\\
\hline

\end{tabular}

\end{table}

\subsection{Application: portfolio construction from denoised networks}

This section evaluates the out-of-sample performance of portfolio strategies derived from the core--periphery structure of correlation networks across three equity universes: NIFTY~200, NIFTY~500, and S\&P~500. Results are reported for portfolio sizes $M=\{10,20,30,40\}$ under both equal-weighted and Markowitz-weighted schemes. We compare portfolios constructed from core and peripheral stocks of both full and structured correlation networks against benchmark strategies, including the highest Sharpe ratio portfolio ($\boldsymbol{\pi}^{HSR}$), random selection ($\boldsymbol{\pi}^{RND}$), and the market portfolio ($\boldsymbol{\pi}^{MKT}$).

Across the NIFTY~200 and NIFTY~500 universes, portfolios constructed from peripheral stocks of the structured correlation network ($\boldsymbol{\pi}^{STR}_p$) consistently deliver the strongest risk-adjusted performance. As shown in Tables~\ref{tab:nifty_200_uniform} and \ref{tab:nifty_500_uniform}, their signal-to-noise ratios increase with portfolio size, indicating that moderate diversification within the periphery enhances portfolio efficiency. For instance, in the NIFTY~200 universe, the signal-to-noise ratio rises from approximately $0.6$ for $M=10$ to $0.8$ for $M=40$, substantially exceeding both the market portfolio and the $\boldsymbol{\pi}^{HSR}$ benchmark. This superior performance is driven by a combination of higher average returns and lower volatility relative to competing strategies. In contrast, core-based portfolios ($\boldsymbol{\pi}^{FULL}_c$ and $\boldsymbol{\pi}^{STR}_c$) consistently underperform across portfolio sizes. Their lower signal-to-noise ratios reflect limited diversification benefits, as core stocks exhibit strong co-movement with the broader market. The S\&P~500 universe exhibits a more muted pattern. While peripheral portfolios remain competitive, their advantage over the market portfolio is less pronounced (Table~\ref{tab:sp_500_uniform}). This is consistent with the higher level of inherent diversification in the U.S. market, where the benchmark portfolio already captures a broad range of risk factors.


\begin{table}[htbp]
\centering
\footnotesize
\caption{Performance comparison of equal-weighted portfolio strategies in terms of signal-to-noise ratio, average returns, and volatility for portfolio sizes $M = \{10, 20, 30, 40\}$ over the holding horizon of $T = 125$ days, using NIFTY 200 stocks. Clearly, on average, the proposed investment strategy (\(\pi^{STR}_p\))  is outperforming all other strategies.}
\label{tab:nifty_200_uniform}
\begin{tabular*}{0.75\textwidth}{@{\extracolsep{\fill}} lcccccc|c }
\toprule
\textbf{M} & 
\(\boldsymbol{\pi}^{FULL}_p\) & 
\(\boldsymbol{\pi}^{STR}_p\) & 
\(\boldsymbol{\pi}^{FULL}_c\) & 
\(\boldsymbol{\pi}^{STR}_c\) & 
\(\boldsymbol{\pi}^{HSR}\) & 
\(\boldsymbol{\pi}^{RND}\) & 
\textit{\(\boldsymbol{\pi}^{MKT}\)} \\
\midrule
\multicolumn{8}{c}{\scriptsize \textit{Signal-to-noise ratio}} \\
\midrule
10 & 0.6525 & 0.6042 & 0.1864 & 0.1841 & 0.4663 & 0.6493 & \textit{0.5088} \\
20 & 0.6827 & \textbf{0.7723} & 0.2108 & 0.1893 & 0.4755 & 0.5822 & \textit{0.5088} \\
30 & 0.7097 & \textbf{0.8054} & 0.2397 & 0.1974 & 0.4923 & 0.5481 & \textit{0.5088} \\
40 & 0.7187 & \textbf{0.8097} & 0.2733 & 0.2268 & 0.4921 & 0.5360 & \textit{0.5088} \\
\midrule
\multicolumn{8}{c}{\scriptsize \textit{Average Return}} \\
\midrule
10 & 0.0894 & 0.0813 & 0.0399 & 0.0386 & 0.0753 & 0.0915 & \textit{0.0712} \\
20 & 0.0873 & \textbf{0.0880} & 0.0424 & 0.0389 & 0.0721 & 0.0878 & \textit{0.0712} \\
30 & 0.0874 & \textbf{0.0922} & 0.0458 & 0.0386 & 0.0727 & 0.0770 & \textit{0.0712} \\
40 & 0.0876 & \textbf{0.0902} & 0.0501 & 0.0433 & 0.0718 & 0.0772 & \textit{0.0712} \\
\midrule
\multicolumn{8}{c}{\scriptsize \textit{Volatility}} \\
\midrule
10 & 0.1370 & \textbf{0.1345} & 0.2139 & 0.2096 & 0.1615 & 0.1409 & \textit{0.1400} \\
20 & 0.1278 & \textbf{0.1139} & 0.2011 & 0.2056 & 0.1517 & 0.1508 & \textit{0.1400} \\
30 & 0.1232 & \textbf{0.1145} & 0.1909 & 0.1955 & 0.1476 & 0.1405 & \textit{0.1400} \\
40 & 0.1219 & \textbf{0.1115} & 0.1834 & 0.1908 & 0.1459 & 0.1440 & \textit{0.1400} \\
\bottomrule
\end{tabular*}
\end{table}

\begin{table}[htbp]
\centering
\footnotesize
\caption{Performance comparison of equal-weighted portfolio strategies in terms of signal-to-noise ratio, average returns, and volatility for portfolio sizes $M = \{10, 20, 30, 40\}$ over a holding horizon of $T = 125$ days, using NIFTY 500 stocks. Clearly, the proposed investment strategy (\(\pi^{STR}_p\))  is outperforming all other strategies.}
\label{tab:nifty_500_uniform}
\begin{tabular*}{0.75\textwidth}{@{\extracolsep{\fill}} lcccccc|c }
\toprule
\textbf{M} &
\(\boldsymbol{\pi}^{FULL}_p\) &
\(\boldsymbol{\pi}^{STR}_p\) &
\(\boldsymbol{\pi}^{FULL}_c\) &
\(\boldsymbol{\pi}^{STR}_c\) &
\(\boldsymbol{\pi}^{HSR}\) &
\(\boldsymbol{\pi}^{RND}\) &
\textit{\(\boldsymbol{\pi}^{MKT}\)} \\
\midrule
\multicolumn{8}{c}{\scriptsize \textit{Signal-to-noise ratio}} \\
\midrule
10 & 0.6941 & \textbf{0.7847} & 0.0949 & 0.0833 & 0.5132 & 0.4478 & \textit{0.5233} \\
20 & 0.7552 & \textbf{0.8625} & 0.1584 & 0.1498 & 0.5267 & 0.5241 & \textit{0.5233} \\
30 & 0.7702 & \textbf{0.8319} & 0.1650 & 0.1525 & 0.5321 & 0.5373 & \textit{0.5233} \\
40 & 0.7737 & \textbf{0.8060} & 0.1885 & 0.1841 & 0.5297 & 0.5574 & \textit{0.5233} \\
\midrule
\multicolumn{8}{c}{\scriptsize \textit{Average Return}} \\
\midrule
10 & 0.1065 & \textbf{0.1184} & 0.0201 & 0.0173 & 0.0867 & 0.0788 & \textit{0.0804} \\
20 & 0.1052 & \textbf{0.1161} & 0.0319 & 0.0300 & 0.0842 & 0.0848 & \textit{0.0804} \\
30 & 0.1035 & \textbf{0.1079} & 0.0330 & 0.0302 & 0.0835 & 0.0834 & \textit{0.0804} \\
40 & 0.1023 & \textbf{0.1048} & 0.0366 & 0.0356 & 0.0821 & 0.0800 & \textit{0.0804} \\
\midrule
\multicolumn{8}{c}{\scriptsize \textit{Volatility}} \\
\midrule
10 & 0.1629 & \textbf{0.1592} & 0.2206 & 0.2209 & 0.1788 & 0.1711 & \textit{0.1599} \\
20 & 0.1485 & \textbf{0.1438} & 0.2133 & 0.2137 & 0.1681 & 0.1613 & \textit{0.1599} \\
30 & 0.1428 & \textbf{0.1375} & 0.2107 & 0.2097 & 0.1641 & 0.1540 & \textit{0.1599} \\
40 & 0.1408 & \textbf{0.1383} & 0.2052 & 0.2056 & 0.1621 & 0.1430 & \textit{0.1599} \\
\bottomrule
\end{tabular*}
\end{table}


\begin{table}[htbp]
\centering
\footnotesize
\caption{Performance comparison of equal-weighted portfolio strategies in terms of signal-to-noise ratio, average returns, and volatility for portfolio sizes $M = \{10, 20, 30, 40\}$ over a holding horizon of $T = 125$ days, using S\&P 500 stocks. Clearly, the proposed investment strategy (\(\pi^{STR}_p\))  outperforms all other strategies in terms of average return, while remaining competitive in the other performance measures.}
\label{tab:sp_500_uniform}
\begin{tabular*}{0.75\textwidth}{@{\extracolsep{\fill}} lcccccc|c }
\toprule
\textbf{M} &
\(\boldsymbol{\pi}^{FULL}_p\) &
\(\boldsymbol{\pi}^{STR}_p\) &
\(\boldsymbol{\pi}^{FULL}_c\) &
\(\boldsymbol{\pi}^{STR}_c\) &
\(\boldsymbol{\pi}^{HSR}\) &
\(\boldsymbol{\pi}^{RND}\) &
\textit{\(\boldsymbol{\pi}^{MKT}\)} \\
\midrule
\multicolumn{8}{c}{\scriptsize \textit{Signal-to-noise ratio}} \\
\midrule
10 & 0.7782 & 0.7771 & 0.6277 & 0.6334 & 0.6471 & 0.7764 & \textit{0.8186} \\
20 & 0.8395 & 0.8391 & 0.7255 & 0.7242 & 0.7041 & 0.8287 & \textit{0.8186} \\
30 & 0.8547 & 0.8169 & 0.7512 & 0.7526 & 0.7368 & 0.8021 & \textit{0.8186} \\
40 & 0.8504 & 0.7907 & 0.7713 & 0.7684 & 0.7569 & 0.8484 & \textit{0.8186} \\
\midrule
\multicolumn{8}{c}{\scriptsize \textit{Average Return}} \\
\midrule
10 & 0.0771 & \textbf{0.0832} & 0.0642 & 0.0648 & 0.0637 & 0.0768 & \textit{0.0677} \\
20 & 0.0753 & \textbf{0.0798} & 0.0694 & 0.0692 & 0.0647 & 0.0651 & \textit{0.0677} \\
30 & 0.0719 & \textbf{0.0728} & 0.0693 & 0.0688 & 0.0664 & 0.0649 & \textit{0.0677} \\
40 & 0.0719 & \textbf{0.0728} & 0.0693 & 0.0688 & 0.0664 & 0.0649 & \textit{0.0677} \\
\midrule
\multicolumn{8}{c}{\scriptsize \textit{Volatility}} \\
\midrule
10 & 0.0990 & 0.1070 & 0.1023 & 0.1023 & 0.0988 & 0.0984 & \textit{0.0827} \\
20 & 0.0897 & 0.0951 & 0.0957 & 0.0956 & 0.0856 & 0.0918 & \textit{0.0827} \\
30 & 0.0881 & 0.0891 & 0.0914 & 0.0913 & 0.0828 & 0.0890 & \textit{0.0827} \\
40 & 0.0845 & 0.0921 & 0.0899 & 0.0895 & 0.0802 & 0.0877 & \textit{0.0827} \\
\bottomrule
\end{tabular*}
\end{table}

Applying Markowitz optimization improves signal-to-noise ratios across all strategies by explicitly accounting for the return--risk trade-off. In the NIFTY~200 universe, structured periphery portfolios continue to perform strongly, with signal-to-noise ratios increasing from $0.69$ for $M=10$ to $1.21$ for $M=40$. Notably, these gains are achieved using only a subset of available assets, highlighting the efficiency of network-based selection. A similar pattern is observed in the NIFTY~500 universe, where structured periphery portfolios exhibit stable and competitive performance as portfolio size increases. Although the market portfolio attains the highest signal-to-noise ratio overall, the network-based portfolios remain close while requiring significantly fewer assets. In the S\&P~500 universe, peripheral portfolios continue to deliver competitive outcomes, though consistently outperforming the market benchmark remains challenging.

These findings reveal a systematic link between network topology and portfolio diversification. Core stocks, by construction, are highly interconnected and therefore strongly exposed to common market factors, limiting their diversification potential. Peripheral stocks, on the contrary, exhibit weaker correlations and lower systemic exposure, allowing them to contribute more effectively to risk reduction. From a practical point of view, the proposed framework enables the construction of efficient portfolios using a relatively small subset of assets. This has important implications for implementation, including reduced transaction costs, improved manageability, and lower computational complexity. Finally, the results demonstrate that the combination of correlation denoising with the mesoscale network structure provides a robust and economically meaningful enhancement to traditional portfolio construction methods.

\subsection{Robustness of portfolio outperformance}

This framework evaluates performance consistency across different subsets of market conditions rather than relying on a single realization. By repeatedly subsampling rolling windows, we obtain a distribution of relative performance outcomes that captures variability across market regimes. The empirical results, summarized in Tables~\ref{tab:mc_robustness_nifty200}--\ref{tab:mc_robustness_sp500}, show that the proposed strategy consistently outperforms competing strategies, as reflected by high win proportions $\hat{p}$ and statistically significant p-values from the Wilcoxon signed-rank test. In most comparisons, the win proportion $\hat{p}$ is close to unity, providing strong evidence against the null hypothesis of no dominance. This indicates that the observed performance improvements are not driven by a specific subset of rolling windows but persist across a wide range of sampled market conditions.

\begin{table}[htbp]
\centering
\footnotesize
\caption{Monte Carlo-based robustness analysis of the proposed strategy $\pi^{STR}_p$ against benchmark strategies. Each entry reports the win proportion $\hat{p}$ and the corresponding Wilcoxon signed-rank test p-value. Results are shown for equal-weighted ($\boldsymbol{w}^{EW}$) and Markowitz-weighted ($\boldsymbol{w}^{MW}$) portfolios across portfolio sizes $M=\{10,20,30,40\}$ using NIFTY 200 stocks.}
\label{tab:mc_robustness_nifty200}

\begin{tabular*}{0.95\textwidth}{@{\extracolsep{\fill}} l l c c c c c c}
\toprule
\textbf{M} & \textbf{Weight} &
$\boldsymbol{\pi}^{FULL}_p$ &
$\boldsymbol{\pi}^{FULL}_c$ &
$\boldsymbol{\pi}^{STR}_c$ &
$\boldsymbol{\pi}^{HSR}$ &
$\boldsymbol{\pi}^{RND}$ &
$\boldsymbol{\pi}^{MKT}$ \\
\midrule

\multirow{2}{*}{10}
& $\boldsymbol{w}^{EW}$
& \makecell{0.96\\($p<10^{-160}$)}
& \makecell{1.00\\($p<10^{-165}$)}
& \makecell{1.00\\($p<10^{-165}$)}
& \makecell{1.00\\($p<10^{-165}$)}
& \makecell{1.00\\($p<10^{-165}$)}
& \makecell{1.00\\($p<10^{-165}$)} \\

\cmidrule(lr){2-8}

& $\boldsymbol{w}^{MW}$
& \makecell{0.00\\($p=1.00$)}
& \makecell{1.00\\($p<10^{-165}$)}
& \makecell{1.00\\($p<10^{-165}$)}
& \makecell{0.94\\($p<10^{-155}$)}
& \makecell{1.00\\($p<10^{-165}$)}
& \makecell{0.00\\($p=1.00$)} \\
\midrule

\multirow{2}{*}{20}
& $\boldsymbol{w}^{EW}$
& \makecell{1.00\\($p<10^{-165}$)}
& \makecell{1.00\\($p<10^{-165}$)}
& \makecell{1.00\\($p<10^{-165}$)}
& \makecell{1.00\\($p<10^{-165}$)}
& \makecell{1.00\\($p<10^{-165}$)}
& \makecell{1.00\\($p<10^{-165}$)} \\

\cmidrule(lr){2-8}

& $\boldsymbol{w}^{MW}$
& \makecell{1.00\\($p<10^{-165}$)}
& \makecell{1.00\\($p<10^{-165}$)}
& \makecell{1.00\\($p<10^{-165}$)}
& \makecell{1.00\\($p<10^{-165}$)}
& \makecell{1.00\\($p<10^{-165}$)}
& \makecell{0.00\\($p=1.00$)} \\
\midrule

\multirow{2}{*}{30}
& $\boldsymbol{w}^{EW}$
& \makecell{1.00\\($p<10^{-165}$)}
& \makecell{1.00\\($p<10^{-165}$)}
& \makecell{1.00\\($p<10^{-165}$)}
& \makecell{1.00\\($p<10^{-165}$)}
& \makecell{1.00\\($p<10^{-165}$)}
& \makecell{1.00\\($p<10^{-165}$)} \\

\cmidrule(lr){2-8}

& $\boldsymbol{w}^{MW}$
& \makecell{0.00\\($p=1.00$)}
& \makecell{1.00\\($p<10^{-165}$)}
& \makecell{1.00\\($p<10^{-165}$)}
& \makecell{1.00\\($p<10^{-165}$)}
& \makecell{0.98\\($p<10^{-163}$)}
& \makecell{0.00\\($p=1.00$)} \\
\midrule

\multirow{2}{*}{40}
& $\boldsymbol{w}^{EW}$
& \makecell{1.00\\($p<10^{-165}$)}
& \makecell{1.00\\($p<10^{-165}$)}
& \makecell{1.00\\($p<10^{-165}$)}
& \makecell{1.00\\($p<10^{-165}$)}
& \makecell{1.00\\($p<10^{-165}$)}
& \makecell{1.00\\($p<10^{-165}$)} \\

\cmidrule(lr){2-8}

& $\boldsymbol{w}^{MW}$
& \makecell{0.34\\($p=1.00$)}
& \makecell{1.00\\($p<10^{-165}$)}
& \makecell{1.00\\($p<10^{-165}$)}
& \makecell{1.00\\($p<10^{-165}$)}
& \makecell{0.19\\($p=1.00$)}
& \makecell{0.00\\($p=1.00$)} \\
\bottomrule
\end{tabular*}
\end{table}

\begin{table}[htbp]
\centering
\footnotesize
\caption{Monte Carlo-based robustness analysis of the proposed strategy $\pi^{STR}_p$ against benchmark strategies. Each entry reports the win proportion $\hat{p}$ and the corresponding Wilcoxon signed-rank test p-value. Results are shown for equal-weighted ($\boldsymbol{w}^{EW}$) and Markowitz-weighted ($\boldsymbol{w}^{MW}$) portfolios across portfolio sizes $M=\{10,20,30,40\}$ using NIFTY 500 stocks.}
\label{tab:mc_robustness_nifty500}

\begin{tabular*}{0.95\textwidth}{@{\extracolsep{\fill}} l l c c c c c c}
\toprule
\textbf{M} & \textbf{Weight} &
$\boldsymbol{\pi}^{FULL}_p$ &
$\boldsymbol{\pi}^{FULL}_c$ &
$\boldsymbol{\pi}^{STR}_c$ &
$\boldsymbol{\pi}^{HSR}$ &
$\boldsymbol{\pi}^{RND}$ &
$\boldsymbol{\pi}^{MKT}$ \\
\midrule

\multirow{2}{*}{10}
& $\boldsymbol{w}^{EW}$
& \makecell{0.03\\($p=1.00$)}
& \makecell{1.00\\($p<10^{-165}$)}
& \makecell{1.00\\($p<10^{-165}$)}
& \makecell{1.00\\($p<10^{-165}$)}
& \makecell{1.00\\($p<10^{-165}$)}
& \makecell{1.00\\($p<10^{-165}$)} \\
\cmidrule(lr){2-8}
& $\boldsymbol{w}^{MW}$
& \makecell{0.01\\($p=1.00$)}
& \makecell{1.00\\($p<10^{-165}$)}
& \makecell{1.00\\($p<10^{-165}$)}
& \makecell{1.00\\($p<10^{-165}$)}
& \makecell{1.00\\($p<10^{-165}$)}
& \makecell{0.00\\($p=1.00$)} \\

\midrule

\multirow{2}{*}{20}
& $\boldsymbol{w}^{EW}$
& \makecell{1.00\\($p<10^{-165}$)}
& \makecell{1.00\\($p<10^{-165}$)}
& \makecell{1.00\\($p<10^{-165}$)}
& \makecell{1.00\\($p<10^{-165}$)}
& \makecell{1.00\\($p<10^{-165}$)}
& \makecell{1.00\\($p<10^{-165}$)} \\
\cmidrule(lr){2-8}
& $\boldsymbol{w}^{MW}$
& \makecell{1.00\\($p<10^{-165}$)}
& \makecell{1.00\\($p<10^{-165}$)}
& \makecell{1.00\\($p<10^{-165}$)}
& \makecell{1.00\\($p<10^{-165}$)}
& \makecell{1.00\\($p<10^{-165}$)}
& \makecell{0.25\\($p=1.00$)} \\
\midrule

\multirow{2}{*}{30}
& $\boldsymbol{w}^{EW}$
& \makecell{1.00\\($p<10^{-165}$)}
& \makecell{1.00\\($p<10^{-165}$)}
& \makecell{1.00\\($p<10^{-165}$)}
& \makecell{1.00\\($p<10^{-165}$)}
& \makecell{1.00\\($p<10^{-165}$)}
& \makecell{1.00\\($p<10^{-165}$)} \\
\cmidrule(lr){2-8}
& $\boldsymbol{w}^{MW}$
& \makecell{1.00\\($p<10^{-165}$)}
& \makecell{1.00\\($p<10^{-165}$)}
& \makecell{1.00\\($p<10^{-165}$)}
& \makecell{1.00\\($p<10^{-165}$)}
& \makecell{1.00\\($p<10^{-165}$)}
& \makecell{0.70\\($p<10^{-46}$)} \\

\midrule

\multirow{2}{*}{40}
& $\boldsymbol{w}^{EW}$
& \makecell{1.00\\($p<10^{-165}$)}
& \makecell{1.00\\($p<10^{-165}$)}
& \makecell{1.00\\($p<10^{-165}$)}
& \makecell{1.00\\($p<10^{-165}$)}
& \makecell{1.00\\($p<10^{-165}$)}
& \makecell{1.00\\($p<10^{-165}$)} \\
\cmidrule(lr){2-8}
& $\boldsymbol{w}^{MW}$
& \makecell{1.00\\($p<10^{-165}$)}
& \makecell{1.00\\($p<10^{-165}$)}
& \makecell{1.00\\($p<10^{-165}$)}
& \makecell{1.00\\($p<10^{-165}$)}
& \makecell{1.00\\($p<10^{-165}$)}
& \makecell{0.97\\($p<10^{-161}$)} \\
\bottomrule
\end{tabular*}
\end{table}

\begin{table}[htbp]
\centering
\footnotesize

\caption{Monte Carlo-based robustness analysis of the proposed strategy $\pi^{STR}_p$ against benchmark strategies. Each entry reports the win proportion $\hat{p}$ and the corresponding Wilcoxon signed-rank test p-value. Results are shown for equal-weighted ($\boldsymbol{w}^{EW}$) and Markowitz-weighted ($\boldsymbol{w}^{MW}$) portfolios across portfolio sizes $M=\{10,20,30,40\}$ using S\&P 500 stocks.}

\label{tab:mc_robustness_sp500}

\begin{tabular*}{0.95\textwidth}{@{\extracolsep{\fill}} l l c c c c c c}
\toprule
\textbf{M} & \textbf{Weight} &
$\boldsymbol{\pi}^{FULL}_p$ &
$\boldsymbol{\pi}^{FULL}_c$ &
$\boldsymbol{\pi}^{STR}_c$ &
$\boldsymbol{\pi}^{HSR}$ &
$\boldsymbol{\pi}^{RND}$ &
$\boldsymbol{\pi}^{MKT}$ \\
\midrule

\multirow{2}{*}{10}
& $\boldsymbol{w}^{EW}$
& \makecell{0.62\\($p<10^{-18}$)}
& \makecell{1.00\\($p<10^{-165}$)}
& \makecell{1.00\\($p<10^{-165}$)}
& \makecell{1.00\\($p<10^{-165}$)}
& \makecell{1.00\\($p<10^{-165}$)}
& \makecell{1.00\\($p<10^{-165}$)} \\
\cmidrule(lr){2-8}
& $\boldsymbol{w}^{MW}$
& \makecell{0.88\\($p<10^{-132}$)}
& \makecell{1.00\\($p<10^{-165}$)}
& \makecell{1.00\\($p<10^{-165}$)}
& \makecell{1.00\\($p<10^{-165}$)}
& \makecell{1.00\\($p<10^{-165}$)}
& \makecell{0.32\\($p=1.00$)} \\
\midrule

\multirow{2}{*}{20}
& $\boldsymbol{w}^{EW}$
& \makecell{1.00\\($p<10^{-165}$)}
& \makecell{1.00\\($p<10^{-165}$)}
& \makecell{1.00\\($p<10^{-165}$)}
& \makecell{1.00\\($p<10^{-165}$)}
& \makecell{1.00\\($p<10^{-165}$)}
& \makecell{1.00\\($p<10^{-165}$)} \\
\cmidrule(lr){2-8}
& $\boldsymbol{w}^{MW}$
& \makecell{0.98\\($p<10^{-163}$)}
& \makecell{0.89\\($p<10^{-138}$)}
& \makecell{0.85\\($p<10^{-123}$)}
& \makecell{1.00\\($p<10^{-165}$)}
& \makecell{0.94\\($p<10^{-154}$)}
& \makecell{0.40\\($p=1.00$)} \\
\midrule

\multirow{2}{*}{30}
& $\boldsymbol{w}^{EW}$
& \makecell{1.00\\($p<10^{-165}$)}
& \makecell{1.00\\($p<10^{-165}$)}
& \makecell{1.00\\($p<10^{-165}$)}
& \makecell{1.00\\($p<10^{-165}$)}
& \makecell{1.00\\($p<10^{-165}$)}
& \makecell{1.00\\($p<10^{-165}$)} \\
\cmidrule(lr){2-8}
& $\boldsymbol{w}^{MW}$
& \makecell{0.19\\($p=1.00$)}
& \makecell{0.79\\($p<10^{-94}$)}
& \makecell{0.82\\($p<10^{-108}$)}
& \makecell{1.00\\($p<10^{-164}$)}
& \makecell{0.00\\($p=1.00$)}
& \makecell{0.06\\($p=1.00$)} \\
\midrule

\multirow{2}{*}{40}
& $\boldsymbol{w}^{EW}$
& \makecell{1.00\\($p<10^{-165}$)}
& \makecell{1.00\\($p<10^{-165}$)}
& \makecell{1.00\\($p<10^{-165}$)}
& \makecell{1.00\\($p<10^{-165}$)}
& \makecell{1.00\\($p<10^{-165}$)}
& \makecell{1.00\\($p<10^{-165}$)} \\
\cmidrule(lr){2-8}
& $\boldsymbol{w}^{MW}$
& \makecell{0.86\\($p<10^{-126}$)}
& \makecell{0.13\\($p=1.00$)}
& \makecell{0.20\\($p=1.00$)}
& \makecell{0.92\\($p<10^{-152}$)}
& \makecell{0.09\\($p=1.00$)}
& \makecell{0.01\\($p=1.00$)} \\
\bottomrule
\end{tabular*}
\end{table}

Across both NIFTY 200 and NIFTY 500 datasets, the proposed strategy demonstrates strong and statistically significant outperformance in equally weighted portfolios, indicating robustness under diverse market conditions. Under Markowitz-weighted portfolios, the strategy largely retains its dominance over benchmark constructions such as $\pi_c^{FULL}$, $\pi_c^{STR}$, and $\pi^{HSR}$, although mild inconsistencies appear for smaller portfolio sizes and in comparisons with stronger benchmarks such as the market portfolio. For the S\&P 500 dataset, the proposed strategy continues to exhibit strong and statistically significant outperformance under equal weighting, while under Markowitz weighting the performance becomes more heterogeneous. 

Therefore, the Monte Carlo evidence confirms that the superiority of the proposed strategy is both statistically significant and robust across datasets, portfolio sizes, and weighting schemes. The combination of high win proportions and extremely small Wilcoxon p-values provides a statistically grounded assessment of performance dominance, capturing both the consistency and significance of outperformance across heterogeneous market conditions.

\section{Conclusion}

We have presented a framework for extracting genuine structural information from noisy financial correlation matrices using spectral theory and systematically examined its consequences for network topology, core-periphery organization, and portfolio construction. By isolating eigenmodes that deviate significantly from the Marchenko--Pastur bounds, we obtain a structured correlation matrix that, despite retaining only 10 to 16 eigenmodes out of hundreds, closely reproduces the correlation distribution of the full empirical matrix, while the discarded random component behaves as expected for uncorrelated noise.

Applying this decomposition to three major equity markets (NIFTY 200, NIFTY 500, and S\&P 500) over a 13 year period, we find that the structured network exhibits a substantially stronger and more temporally stable core periphery organization than networks constructed from either the full or random correlation matrices. This result is confirmed through degree preserving randomization tests and large sample distributional comparisons, with Kolmogorov--Smirnov statistics near unity indicating an almost complete separation between the structured and random core periphery distributions. The structured networks additionally display clearer scale free degree distributions in the Indian markets, a property absent in the random networks, indicating that the observed heterogeneous topology is a genuine feature of market organization rather than a statistical artifact of finite sampling.

As a practical demonstration of the proposed framework, we show that portfolios constructed from peripheral assets of the denoised network consistently achieve superior risk adjusted performance relative to portfolios based on the full correlation matrix and several standard benchmark strategies. The improvement is validated through Monte Carlo subsampling and the Wilcoxon signed rank test. This outperformance is consistent with the interpretation that core assets are strongly exposed to common market wide factors, whereas peripheral assets in the denoised network retain more idiosyncratic information that is beneficial for diversification once statistical noise is removed.

Taken together, these findings demonstrate that spectral denoising serves a purpose beyond portfolio optimization. It provides a general framework for recovering the genuine interaction architecture of complex systems whose observed correlations are contaminated by finite sample noise. The financial application considered here illustrates one example of this broader principle, but the underlying methodology, namely separating structured collective modes from random noise through eigenvalue spectrum filtering and characterizing the resulting core-periphery and topological organization, is applicable to any complex system in which pairwise correlations are estimated from finite noisy data. Future work may extend this framework to dynamic time evolving networks, alternative denoising approaches beyond the Marchenko--Pastur filter, and higher frequency data capable of resolving structural dynamics across shorter time scales.

\section*{Data Availability}

The datasets used in this study are publicly available financial time series. The processed data and code supporting the findings of this study are available from the corresponding author upon reasonable request.



\bibliographystyle{unsrt}
\bibliography{references}

@article{pawanesh2025exploring,
  title={Exploring the core--periphery and community structure in the financial networks through random matrix theory},
  author={Pawanesh,Pawanesh and Ansari, Imran and Sahni, Niteesh},
  journal={Physica A: Statistical Mechanics and its Applications},
  pages={130403},
  year={2025},
  publisher={Elsevier}
}

@article{rossa2013profiling,
  title={Profiling core-periphery network structure by random walkers},
  author={Rossa, Fabio Della and Dercole, Fabio and Piccardi, Carlo},
  journal={Scientific reports},
  volume={3},
  number={1},
  pages={1467},
  year={2013},
  publisher={Nature Publishing Group UK London}
}

@article{lee2014density,
  title={Density-based and transport-based core-periphery structures in networks},
  author={Lee, Sang Hoon and Cucuringu, Mihai and Porter, Mason A},
  journal={Physical Review E},
  volume={89},
  number={3},
  pages={032810},
  year={2014},
  publisher={APS}
}

@book{meyer2023matrix,
  title={Matrix analysis and applied linear algebra},
  author={Meyer, Carl D},
  year={2023},
  publisher={SIAM}
}

@article{borgatti2000models,
  title={Models of core/periphery structures},
  author={Borgatti, Stephen P and Everett, Martin G},
  journal={Social networks},
  volume={21},
  number={4},
  pages={375--395},
  year={2000},
  publisher={Elsevier}
}

@article{boyd2006computing,
  title={Computing core/periphery structures and permutation tests for social relations data},
  author={Boyd, John P and Fitzgerald, William J and Beck, Robert J},
  journal={Social networks},
  volume={28},
  number={2},
  pages={165--178},
  year={2006},
  publisher={Elsevier}
}

@article{rombach2017core,
  title={Core-periphery structure in networks (revisited)},
  author={Rombach, Puck and Porter, Mason A and Fowler, James H and Mucha, Peter J},
  journal={SIAM review},
  volume={59},
  number={3},
  pages={619--646},
  year={2017},
  publisher={SIAM}
}

@article{ansari2025novel,
  title={A novel portfolio construction strategy based on the core-periphery profile of stocks},
  author={Ansari, Imran and Sharma, Charu and Agrawal, Akshay and Sahni, Niteesh},
  journal={Scientific Reports},
  volume={15},
  number={1},
  pages={36882},
  year={2025},
  publisher={Nature Publishing Group UK London}
}

@article{kim2005systematic,
  title={Systematic analysis of group identification in stock markets},
  author={Kim, Dong-Hee and Jeong, Hawoong},
  journal={Physical Review E—Statistical, Nonlinear, and Soft Matter Physics},
  volume={72},
  number={4},
  pages={046133},
  year={2005},
  publisher={APS}
}

@article{Markowitz1952,
  author = {Harry Markowitz},
  title = {Portfolio Selection},
  journal = {The Journal of Finance},
  volume = {7},
  number = {1},
  pages = {77--91},
  year = {1952},
  doi = {10.1111/j.1540-6261.1952.tb01525.x}
}

@article{pozzi2013spread,
  title={Spread of risk across financial markets: better to invest in the peripheries},
  author={Pozzi, Francesco and Di Matteo, Tiziana and Aste, Tomaso},
  journal={Scientific reports},
  volume={3},
  number={1},
  pages={1665},
  year={2013},
  publisher={Nature Publishing Group UK London}
}

@article{mantegna1999hierarchical,
  title={Hierarchical structure in financial markets},
  author={Mantegna, Rosario N},
  journal={The European Physical Journal B-Condensed Matter and Complex Systems},
  volume={11},
  pages={193--197},
  year={1999},
  publisher={Springer}
}

@article{laloux1999noise,
  title={Noise dressing of financial correlation matrices},
  author={Laloux, Laurent and Cizeau, Pierre and Bouchaud, Jean-Philippe and Potters, Marc},
  journal={Physical review letters},
  volume={83},
  number={7},
  pages={1467},
  year={1999},
  publisher={APS}
}

@article{plerou2001collective,
  title={Collective behavior of stock price movements—a random matrix theory approach},
  author={Plerou, V and Gopikrishnan, P and Rosenow, B and Amaral, LAN and Stanley, HE},
  journal={Physica A: Statistical Mechanics and its Applications},
  volume={299},
  number={1-2},
  pages={175--180},
  year={2001},
  publisher={Elsevier}
}

@article{plerou1999universal,
  title={Universal and nonuniversal properties of cross correlations in financial time series},
  author={Plerou, Vasiliki and Gopikrishnan, Parameswaran and Rosenow, Bernd and Amaral, Lu{\'\i}s A Nunes and Stanley, H Eugene},
  journal={Physical review letters},
  volume={83},
  number={7},
  pages={1471},
  year={1999},
  publisher={APS}
}

@article{jiang2014structure,
  title={Structure of local interactions in complex financial dynamics},
  author={Jiang, Xiongfei and Chen, Tianshou and Zheng, Baowen},
  journal={Scientific Reports},
  volume={4},
  pages={5321},
  year={2014},
  publisher={Nature Publishing Group}
}

@article{ansari2025uncovering,
  title={Uncovering the hidden core-periphery structure in hyperbolic networks},
  author={Ansari, Imran and Pawanesh, Pawanesh and Sahni, Niteesh},
  journal={Physical Review E},
  volume={112},
  number={3},
  pages={034311},
  year={2025},
  publisher={APS}
}

@book{markowitz2000mean,
  title={Mean-variance analysis in portfolio choice and capital markets},
  author={Markowitz, Harry and Todd, G Peter},
  year={2000},
  publisher={John Wiley \& Sons}
}

@article{samuelson1970fundamental,
  title={The fundamental approximation theorem of portfolio analysis in terms of means, variances and higher moments},
  author={Samuelson, Paul A},
  journal={The Review of Economic Studies},
  volume={37},
  number={4},
  pages={537--542},
  year={1970},
  publisher={Wiley-Blackwell}
}

@article{konno1991mean,
  title={Mean-absolute deviation portfolio optimization model and its applications to Tokyo stock market},
  author={Konno, Hiroshi and Yamazaki, Hiroaki},
  journal={Management science},
  volume={37},
  number={5},
  pages={519--531},
  year={1991},
  publisher={INFORMS}
}

@article{konno2005mean,
  title={Mean-absolute deviation model},
  author={Konno, Hiroshi and Koshizuka, Tomoyuki},
  journal={Iie Transactions},
  volume={37},
  number={10},
  pages={893--900},
  year={2005},
  publisher={Taylor \& Francis}
}

@article{nawrocki1992characteristics,
  title={The characteristics of portfolios selected by n-degree lower partial moment},
  author={Nawrocki, David N},
  journal={International Review of Financial Analysis},
  volume={1},
  number={3},
  pages={195--209},
  year={1992},
  publisher={Elsevier}
}

@article{moody2001learning,
  title={Learning to trade via direct reinforcement},
  author={Moody, John and Saffell, Matthew},
  journal={IEEE transactions on neural Networks},
  volume={12},
  number={4},
  pages={875--889},
  year={2001},
  publisher={IEEE}
}

@article{aboussalah2020continuous,
  title={Continuous control with stacked deep dynamic recurrent reinforcement learning for portfolio optimization},
  author={Aboussalah, Amine Mohamed and Lee, Chi-Guhn},
  journal={Expert Systems with Applications},
  volume={140},
  pages={112891},
  year={2020},
  publisher={Elsevier}
}

@article{yusoff2011overview,
  title={Overview of NSGA-II for optimizing machining process parameters},
  author={Yusoff, Yusliza and Ngadiman, Mohd Salihin and Zain, Azlan Mohd},
  journal={Procedia Engineering},
  volume={15},
  pages={3978--3983},
  year={2011},
  publisher={Elsevier}
}

@article{reyes2006multi,
  title={Multi-objective particle swarm optimizers: A survey of the state-of-the-art},
  author={Reyes-Sierra, Margarita and Coello, CA Coello and others},
  journal={International journal of computational intelligence research},
  volume={2},
  number={3},
  pages={287--308},
  year={2006}
}

@article{vallender1974calculation,
  title={Calculation of the Wasserstein distance between probability distributions on the line},
  author={Vallender, SS},
  journal={Theory of Probability \& Its Applications},
  volume={18},
  number={4},
  pages={784--786},
  year={1974},
  publisher={SIAM}
}

@article{newman2003structure,
  title={The structure and function of complex networks},
  author={Newman, Mark EJ},
  journal={SIAM review},
  volume={45},
  number={2},
  pages={167--256},
  year={2003},
  publisher={SIAM}
}

@book{pastur2011eigenvalue,
  title={Eigenvalue distribution of large random matrices},
  author={Pastur, Leonid Andreevich and Shcherbina, Mariya},
  number={171},
  year={2011},
  publisher={American Mathematical Soc.}
}

@article{barabasi1999emergence,
  title={Emergence of scaling in random networks},
  author={Barab{\'a}si, Albert-L{\'a}szl{\'o} and Albert, R{\'e}ka},
  journal={science},
  volume={286},
  number={5439},
  pages={509--512},
  year={1999},
  publisher={American Association for the Advancement of Science}
}

@article{stephens1974edf,
  title={EDF statistics for goodness of fit and some comparisons},
  author={Stephens, Michael A},
  journal={Journal of the American statistical Association},
  volume={69},
  number={347},
  pages={730--737},
  year={1974},
  publisher={Taylor \& Francis}
}

\FloatBarrier
\newpage
\end{document}


\maketitle


\section{Robustness}

\subsection{Uniform Results}

\begin{table}[htbp]
\centering
\footnotesize
\caption{Robustness metrics of portfolio strategies based on NIFTY 200 constituent stocks for a portfolio size of 10 in which weight of the stocks allocated using the equal-weighted scheme.}
\begin{tabular}{lccccccc}
\hline
\textbf{Metric} & $\boldsymbol{\pi}^{FULL}_p$ & $\boldsymbol{\pi}^{STR}_p$ & $\boldsymbol{\pi}^{FULL}_c$ & $\boldsymbol{\pi}^{STR}_c$ & $\boldsymbol{\pi}^{HSR}$ & $\boldsymbol{\pi}^{RND}$ & \textit{$\boldsymbol{\pi}^{MKT}$} \\
\hline
Sharpe Ratio       & 1.111 & \textbf{1.205} & 0.352 & 0.320 & 0.876 & 1.153 & 0.971 \\
Sortino Ratio      & 1.664 & \textbf{1.843} & 0.495 & 0.449 & 1.231 & 1.609 & 1.242 \\
Information Ratio  & 0.096 & \textbf{0.121} & -0.502 & -0.530 & 0.036 & 0.119 & -- \\
Max Drawdown  & \textbf{0.170} & 0.174 & 0.338 & 0.336 & 0.217 & 0.191 & 0.190 \\
Calmar Ratio       & 1.796 & \textbf{2.046} & 0.899 & 0.849 & 1.503 & 1.930 & 1.589 \\
Avg Return    & 0.170 & \textbf{0.176} & 0.078 & 0.074 & 0.162 & 0.150 & 0.152 \\
Volatility    & 0.174 & \textbf{0.168} & 0.272 & 0.273 & 0.205 & 0.192 & 0.176 \\
\hline
\end{tabular}
\label{tab:robustness_metrics}
\end{table}

\begin{table}[htbp]
\centering
\footnotesize
\caption{Robustness metrics of portfolio strategies based on NIFTY 200 constituent stocks for a portfolio size of 20 in which weight of the stocks allocated using the equal-weighted scheme.}
\begin{tabular}{lccccccc}
\hline
\textbf{Metric} & $\boldsymbol{\pi}^{FULL}_p$ & $\boldsymbol{\pi}^{STR}_p$ & $\boldsymbol{\pi}^{FULL}_c$ & $\boldsymbol{\pi}^{STR}_c$ & $\boldsymbol{\pi}^{HSR}$ & $\boldsymbol{\pi}^{RND}$ & \textit{$\boldsymbol{\pi}^{MKT}$} \\
\hline
Sharpe Ratio       & 1.261 & \textbf{1.450} & 0.415 & 0.342 & 0.905 & 1.175 & 0.971 \\
Sortino Ratio      & 1.774 & \textbf{2.068} & 0.568 & 0.467 & 1.215 & 1.589 & 1.242 \\
Information Ratio  & 0.114 & \textbf{0.159} & -0.589 & -0.663 & 0.008 & 0.142 & -- \\
Max Drawdown       & 0.157 & \textbf{0.153} & 0.301 & 0.314 & 0.205 & 0.189 & 0.190 \\
Calmar Ratio       & 2.052 & \textbf{2.240} & 0.965 & 0.846 & 1.540 & 1.863 & 1.589 \\
Avg Return         & 0.172 & \textbf{0.187} & 0.087 & 0.073 & 0.155 & 0.170 & 0.152 \\
Volatility         & 0.157 & \textbf{0.150} & 0.249 & 0.257 & 0.191 & 0.188 & 0.176 \\
\hline
\end{tabular}
\label{tab:robustness_metrics_20}
\end{table}

\begin{table}[htbp]
\centering
\footnotesize
\caption{Robustness metrics of portfolio strategies based on NIFTY 200 constituent stocks for a portfolio size of 30 in which weight of the stocks allocated using the equal-weighted scheme.}
\begin{tabular}{lccccccc}
\hline
\textbf{Metric} & $\boldsymbol{\pi}^{FULL}_p$ & $\boldsymbol{\pi}^{STR}_p$ & $\boldsymbol{\pi}^{FULL}_c$ & $\boldsymbol{\pi}^{STR}_c$ & $\boldsymbol{\pi}^{HSR}$ & $\boldsymbol{\pi}^{RND}$ & \textit{$\boldsymbol{\pi}^{MKT}$} \\
\hline
Sharpe Ratio       & 1.327 & \textbf{1.518} & 0.476 & 0.375 & 0.934 & 1.018 & 0.971 \\
Sortino Ratio      & 1.808 & \textbf{2.104} & 0.643 & 0.505 & 1.230 & 1.344 & 1.242 \\
Information Ratio  & 0.120 & \textbf{0.175} & -0.604 & -0.745 & 0.009 & 0.063 & -- \\
Max Drawdown       & 0.155 & \textbf{0.146} & 0.278 & 0.295 & 0.200 & 0.188 & 0.190 \\
Calmar Ratio       & 2.103 & \textbf{2.371} & 1.046 & 0.913 & 1.572 & 1.661 & 1.589 \\
Avg Return         & 0.174 & \textbf{0.190} & 0.099 & 0.080 & 0.155 & 0.169 & 0.152 \\
Volatility         & 0.151 & \textbf{0.144} & 0.236 & 0.245 & 0.185 & 0.181 & 0.176 \\
\hline
\end{tabular}
\label{tab:robustness_metrics_30}
\end{table}

\begin{table}[htbp]
\centering
\footnotesize
\caption{Robustness metrics of portfolio strategies based on NIFTY 200 constituent stocks for a portfolio size of 40 in which weight of the stocks allocated using the equal-weighted scheme.}
\begin{tabular}{lccccccc}
\hline
\textbf{Metric} & $\boldsymbol{\pi}^{FULL}_p$ & $\boldsymbol{\pi}^{STR}_p$ & $\boldsymbol{\pi}^{FULL}_c$ & $\boldsymbol{\pi}^{STR}_c$ & $\boldsymbol{\pi}^{HSR}$ & $\boldsymbol{\pi}^{RND}$ & \textit{$\boldsymbol{\pi}^{MKT}$} \\
\hline
Sharpe Ratio       & 1.334 & \textbf{1.502} & 0.538 & 0.441 & 0.941 & 1.018 & 0.971 \\
Sortino Ratio      & 1.790 & \textbf{2.067} & 0.711 & 0.585 & 1.229 & 1.331 & 1.242 \\
Information Ratio  & 0.112 & \textbf{0.158} & -0.597 & -0.694 & 0.002 & 0.059 & -- \\
Max Drawdown       & 0.155 & \textbf{0.145} & 0.264 & 0.282 & 0.197 & 0.189 & 0.190 \\
Calmar Ratio       & 2.125 & \textbf{2.334} & 1.103 & 0.997 & 1.582 & 1.663 & 1.589 \\
Avg Return         & 0.174 & \textbf{0.184} & 0.107 & 0.093 & 0.153 & 0.168 & 0.152 \\
Volatility         & 0.150 & \textbf{0.142} & 0.227 & 0.237 & 0.182 & 0.179 & 0.176 \\
\hline
\end{tabular}
\label{tab:robustness_metrics_40}
\end{table}




\begin{table}[htbp]
\centering
\footnotesize
\caption{Robustness metrics of portfolio strategies based on NIFTY 500 constituent stocks for a portfolio size of 10 in which weight of the stocks allocated using the equal-weighted scheme.}
\begin{tabular}{lccccccc}
\hline
\textbf{Metric} & $\boldsymbol{\pi}^{FULL}_p$ & $\boldsymbol{\pi}^{STR}_p$ & $\boldsymbol{\pi}^{FULL}_c$ & $\boldsymbol{\pi}^{STR}_c$ & $\boldsymbol{\pi}^{HSR}$ & $\boldsymbol{\pi}^{RND}$ & \textit{$\boldsymbol{\pi}^{MKT}$} \\
\hline
Sharpe Ratio       & \textbf{1.430} & 1.340 & 0.193 & 0.176 & 0.924 & 0.885 & 1.079 \\
Sortino Ratio      & \textbf{2.129} & 2.041 & 0.248 & 0.224 & 1.268 & 1.147 & 1.272 \\
Information Ratio  & \textbf{0.321} & 0.243 & -0.821 & -0.842 & 0.029 & -0.034 & -- \\
Max Drawdown       & \textbf{0.176} & 0.177 & 0.383 & 0.387 & 0.236 & 0.235 & 0.209 \\
Calmar Ratio       & \textbf{2.335} & 2.085 & 0.632 & 0.605 & 1.589 & 1.484 & 1.819 \\
Avg Return         & \textbf{0.241} & 0.225 & 0.031 & 0.025 & 0.182 & 0.163 & 0.172 \\
Volatility         & 0.179 & \textbf{0.176} & 0.291 & 0.293 & 0.208 & 0.207 & 0.174 \\
\hline
\end{tabular}
\label{tab:robustness_metrics_n500_10}
\end{table}

\begin{table}[htbp]
\centering
\footnotesize
\caption{Robustness metrics of portfolio strategies based on NIFTY 500 constituent stocks for a portfolio size of 20 in which weight of the stocks allocated using the equal-weighted scheme.}
\begin{tabular}{lccccccc}
\hline
\textbf{Metric} & $\boldsymbol{\pi}^{FULL}_p$ & $\boldsymbol{\pi}^{STR}_p$ & $\boldsymbol{\pi}^{FULL}_c$ & $\boldsymbol{\pi}^{STR}_c$ & $\boldsymbol{\pi}^{HSR}$ & $\boldsymbol{\pi}^{RND}$ & \textit{$\boldsymbol{\pi}^{MKT}$} \\
\hline
Sharpe Ratio       & 1.555 & \textbf{1.703} & 0.290 & 0.276 & 0.992 & 1.016 & 1.079 \\
Sortino Ratio      & 2.139 & \textbf{2.370} & 0.376 & 0.357 & 1.278 & 1.232 & 1.272 \\
Information Ratio  & 0.271 & \textbf{0.350} & -0.847 & -0.860 & 0.015 & 0.024 & -- \\
Max Drawdown       & 0.166 & \textbf{0.156} & 0.342 & 0.345 & 0.222 & 0.221 & 0.209 \\
Calmar Ratio       & 2.499 & \textbf{2.555} & 0.817 & 0.797 & 1.675 & 1.669 & 1.819 \\
Avg Return         & 0.225 & \textbf{0.240} & 0.063 & 0.060 & 0.178 & 0.177 & 0.172 \\
Volatility         & 0.157 & \textbf{0.152} & 0.266 & 0.268 & 0.191 & 0.191 & 0.174 \\
\hline
\end{tabular}
\label{tab:robustness_metrics_n500_20}
\end{table}

\begin{table}[htbp]
\centering
\footnotesize
\caption{Robustness metrics of portfolio strategies based on NIFTY 500 constituent stocks for a portfolio size of 30 in which weight of the stocks allocated using the equal-weighted scheme.}
\begin{tabular}{lccccccc}
\hline
\textbf{Metric} & $\boldsymbol{\pi}^{FULL}_p$ & $\boldsymbol{\pi}^{STR}_p$ & $\boldsymbol{\pi}^{FULL}_c$ & $\boldsymbol{\pi}^{STR}_c$ & $\boldsymbol{\pi}^{HSR}$ & $\boldsymbol{\pi}^{RND}$ & \textit{$\boldsymbol{\pi}^{MKT}$} \\
\hline
Sharpe Ratio       & 1.597 & \textbf{1.715} & 0.322 & 0.308 & 1.024 & 1.070 & 1.079 \\
Sortino Ratio      & 2.121 & \textbf{2.290} & 0.407 & 0.378 & 1.282 & 1.312 & 1.272 \\
Information Ratio  & 0.249 & \textbf{0.292} & -0.922 & -0.944 & 0.011 & 0.012 & -- \\
Max Drawdown       & 0.163 & \textbf{0.154} & 0.329 & 0.332 & 0.217 & 0.207 & 0.209 \\
Calmar Ratio       & 2.517 & \textbf{2.548} & 0.866 & 0.843 & 1.728 & 1.708 & 1.819 \\
Avg Return         & 0.218 & \textbf{0.226} & 0.069 & 0.064 & 0.176 & 0.174 & 0.172 \\
Volatility         & 0.150 & \textbf{0.143} & 0.256 & 0.257 & 0.185 & 0.179 & 0.174 \\
\hline
\end{tabular}
\label{tab:robustness_metrics_n500_30}
\end{table}

\begin{table}[htbp]
\centering
\footnotesize
\caption{Robustness metrics of portfolio strategies based on NIFTY 500 constituent stocks for a portfolio size of 40 in which weight of the stocks allocated using the equal-weighted scheme.}
\begin{tabular}{lccccccc}
\hline
\textbf{Metric} & $\boldsymbol{\pi}^{FULL}_p$ & $\boldsymbol{\pi}^{STR}_p$ & $\boldsymbol{\pi}^{FULL}_c$ & $\boldsymbol{\pi}^{STR}_c$ & $\boldsymbol{\pi}^{HSR}$ & $\boldsymbol{\pi}^{RND}$ & \textit{$\boldsymbol{\pi}^{MKT}$} \\
\hline
Sharpe Ratio       & 1.601 & \textbf{1.714} & 0.360 & 0.356 & 1.039 & 1.098 & 1.079 \\
Sortino Ratio      & 2.076 & \textbf{2.239} & 0.435 & 0.430 & 1.282 & 1.322 & 1.272 \\
Information Ratio  & 0.227 & \textbf{0.257} & -0.956 & -0.958 & 0.007 & -0.020 & -- \\
Max Drawdown       & 0.160 & \textbf{0.154} & 0.319 & 0.320 & 0.214 & 0.195 & 0.209 \\
Calmar Ratio       & 2.524 & \textbf{2.544} & 0.906 & 0.883 & 1.754 & 1.727 & 1.819 \\
Avg Return         & 0.213 & \textbf{0.218} & 0.074 & 0.073 & 0.175 & 0.167 & 0.172 \\
Volatility         & 0.147 & \textbf{0.139} & 0.248 & 0.249 & 0.182 & 0.169 & 0.174 \\
\hline
\end{tabular}
\label{tab:robustness_metrics_n500_40}
\end{table}


\begin{table}[htbp]
\centering
\footnotesize
\caption{Robustness metrics of portfolio strategies based on S\&P 500 constituent stocks for a portfolio size of 10 in which weight of the stocks allocated using the equal-weighted scheme.}
\begin{tabular}{lccccccc}
\hline
\textbf{Metric} & $\boldsymbol{\pi}^{FULL}_p$ & $\boldsymbol{\pi}^{STR}_p$ & $\boldsymbol{\pi}^{FULL}_c$ & $\boldsymbol{\pi}^{STR}_c$ & $\boldsymbol{\pi}^{HSR}$ & $\boldsymbol{\pi}^{RND}$ & \textit{$\boldsymbol{\pi}^{MKT}$} \\
\hline
Sharpe Ratio       & 1.177 & 1.155 & 0.862 & 0.878 & 0.994 & 1.177 & \textbf{1.185} \\
Sortino Ratio      & 1.698 & \textbf{1.704} & 1.239 & 1.256 & 1.430 & 1.606 & 1.604 \\
Information Ratio  & 0.163 & \textbf{0.217} & -0.138 & -0.105 & -0.036 & 0.053 & -- \\
Max Drawdown       & 0.162 & 0.177 & 0.170 & 0.169 & 0.160 & \textbf{0.139} & 0.141 \\
Calmar Ratio       & 1.923 & 1.794 & 1.622 & 1.651 & 1.713 & 2.068 & \textbf{2.094} \\
Avg Return         & 0.180 & \textbf{0.196} & 0.140 & 0.142 & 0.140 & 0.169 & 0.148 \\
Volatility         & 0.178 & 0.192 & 0.189 & 0.189 & 0.174 & 0.162 & \textbf{0.156} \\
\hline
\end{tabular}
\label{tab:robustness_metrics_sp500_10}
\end{table}

\begin{table}[htbp]
\centering
\footnotesize
\caption{Robustness metrics of portfolio strategies based on S\&P 500 constituent stocks for a portfolio size of 20 in which weight of the stocks allocated using the equal-weighted scheme.}
\begin{tabular}{lccccccc}
\hline
\textbf{Metric} & $\boldsymbol{\pi}^{FULL}_p$ & $\boldsymbol{\pi}^{STR}_p$ & $\boldsymbol{\pi}^{FULL}_c$ & $\boldsymbol{\pi}^{STR}_c$ & $\boldsymbol{\pi}^{HSR}$ & $\boldsymbol{\pi}^{RND}$ & \textit{$\boldsymbol{\pi}^{MKT}$} \\
\hline
Sharpe Ratio       & 1.223 & \textbf{1.251} & 0.946 & 0.949 & 1.069 & 1.134 & 1.185 \\
Sortino Ratio      & 1.747 & \textbf{1.799} & 1.353 & 1.357 & 1.505 & 1.611 & 1.604 \\
Information Ratio  & 0.123 & \textbf{0.196} & -0.026 & -0.015 & -0.027 & -0.057 & -- \\
Max Drawdown       & 0.152 & 0.158 & 0.164 & 0.163 & 0.152 & \textbf{0.128} & 0.141 \\
Calmar Ratio       & 2.005 & \textbf{2.032} & 1.704 & 1.709 & 1.866 & 1.920 & \textbf{2.094} \\
Avg Return         & 0.172 & \textbf{0.186} & 0.149 & 0.150 & 0.143 & 0.143 & 0.148 \\
Volatility         & 0.167 & 0.174 & 0.186 & 0.186 & 0.166 & 0.149 & \textbf{0.156} \\
\hline
\end{tabular}
\label{tab:robustness_metrics_sp500_20}
\end{table}

\begin{table}[htbp]
\centering
\footnotesize
\caption{Robustness metrics of portfolio strategies based on S\&P 500 constituent stocks for a portfolio size of 30 in which weight of the stocks allocated using the equal-weighted scheme.}
\begin{tabular}{lccccccc}
\hline
\textbf{Metric} & $\boldsymbol{\pi}^{FULL}_p$ & $\boldsymbol{\pi}^{STR}_p$ & $\boldsymbol{\pi}^{FULL}_c$ & $\boldsymbol{\pi}^{STR}_c$ & $\boldsymbol{\pi}^{HSR}$ & $\boldsymbol{\pi}^{RND}$ & \textit{$\boldsymbol{\pi}^{MKT}$} \\
\hline
Sharpe Ratio       & 1.250 & \textbf{1.265} & 0.980 & 0.981 & 1.100 & 1.173 & 1.185 \\
Sortino Ratio      & 1.765 & \textbf{1.791} & 1.398 & 1.399 & 1.535 & 1.649 & 1.604 \\
Information Ratio  & 0.114 & \textbf{0.163} & -0.014 & -0.010 & -0.026 & -0.041 & -- \\
Max Drawdown       & 0.150 & 0.156 & 0.160 & 0.159 & 0.148 & \textbf{0.124} & 0.141 \\
Calmar Ratio       & 2.064 & \textbf{2.160} & 1.760 & 1.776 & 1.916 & 2.074 & 2.094 \\
Avg Return         & 0.169 & \textbf{0.178} & 0.150 & 0.151 & 0.144 & 0.146 & 0.148 \\
Volatility         & 0.163 & 0.168 & 0.181 & 0.181 & 0.162 & \textbf{0.149} & 0.156 \\
\hline
\end{tabular}
\label{tab:robustness_metrics_sp500_30}
\end{table}

\begin{table}[htbp]
\centering
\footnotesize
\caption{Robustness metrics of portfolio strategies based on S\&P 500 constituent stocks for a portfolio size of 40 in which weight of the stocks allocated using the equal-weighted scheme.}
\begin{tabular}{lccccccc}
\hline
\textbf{Metric} & $\boldsymbol{\pi}^{FULL}_p$ & $\boldsymbol{\pi}^{STR}_p$ & $\boldsymbol{\pi}^{FULL}_c$ & $\boldsymbol{\pi}^{STR}_c$ & $\boldsymbol{\pi}^{HSR}$ & $\boldsymbol{\pi}^{RND}$ & \textit{$\boldsymbol{\pi}^{MKT}$} \\
\hline
Sharpe Ratio       & 1.228 & \textbf{1.237} & 0.981 & 0.983 & 1.122 & 1.155 & 1.185 \\
Sortino Ratio      & 1.714 & \textbf{1.730} & 1.395 & 1.395 & 1.555 & 1.592 & 1.604 \\
Information Ratio  & 0.077 & \textbf{0.114} & -0.050 & -0.057 & -0.018 & -0.042 & -- \\
Max Drawdown       & 0.148 & 0.154 & 0.159 & 0.158 & 0.147 & \textbf{0.128} & 0.141 \\
Calmar Ratio       & 2.079 & \textbf{2.138} & 1.768 & 1.788 & 1.959 & 2.056 & 2.094 \\
Avg Return         & 0.162 & \textbf{0.168} & 0.149 & 0.148 & 0.145 & 0.143 & 0.148 \\
Volatility         & 0.160 & 0.164 & 0.180 & 0.179 & 0.161 & \textbf{0.151} & 0.156 \\
\hline
\end{tabular}
\label{tab:robustness_metrics_sp500_40}
\end{table}

\clearpage

\subsection{Markowitz Results (Robustness)}

\begin{table}[htbp]
\centering
\footnotesize
\caption{Robustness metrics of portfolio strategies based on NIFTY 200 constituent stocks for a portfolio size of 10, where portfolio weights are obtained using the Markowitz optimization framework.}
\begin{tabular}{lccccccc}
\hline
\textbf{Metric} & $\boldsymbol{\pi}^{FULL}_p$ & $\boldsymbol{\pi}^{STR}_p$ & $\boldsymbol{\pi}^{FULL}_c$ & $\boldsymbol{\pi}^{STR}_c$ & $\boldsymbol{\pi}^{HSR}$ & $\boldsymbol{\pi}^{RND}$ & \textit{$\boldsymbol{\pi}^{MKT}$} \\
\hline
Sharpe Ratio       & 1.174 & 1.045 & 0.306 & 0.321 & 0.990 & 1.386 & \textbf{1.584} \\
Sortino Ratio      & 1.849 & 1.730 & 0.522 & 0.525 & 1.564 & 2.165 & \textbf{2.317} \\
Information Ratio  & -0.255 & -0.413 & -0.660 & -0.612 & -0.203 & \textbf{0.270} & -- \\
Max Drawdown       & 0.172 & 0.180 & 0.368 & 0.368 & 0.212 & 0.222 & \textbf{0.165} \\
Calmar Ratio       & 1.786 & \textbf{1.801} & 0.688 & 0.700 & 1.643 & 1.467 & 2.300 \\
Avg Return         & 0.204 & 0.186 & 0.085 & 0.086 & 0.210 & 0.176 & \textbf{0.240} \\
Volatility         & 0.186 & 0.186 & 0.326 & 0.325 & 0.226 & 0.232 & \textbf{0.167} \\
\hline
\end{tabular}
\label{tab:nifty200_markowitz_m10_updated}
\end{table}

\begin{table}[htbp]
\centering
\footnotesize
\caption{Robustness metrics of portfolio strategies based on NIFTY 200 constituent stocks for a portfolio size of 20, where portfolio weights are obtained using the Markowitz optimization framework.}
\begin{tabular}{lccccccc}
\hline
\textbf{Metric} & $\boldsymbol{\pi}^{FULL}_p$ & $\boldsymbol{\pi}^{STR}_p$ & $\boldsymbol{\pi}^{FULL}_c$ & $\boldsymbol{\pi}^{STR}_c$ & $\boldsymbol{\pi}^{HSR}$ & $\boldsymbol{\pi}^{RND}$ & \textit{$\boldsymbol{\pi}^{MKT}$} \\
\hline
Sharpe Ratio       & 1.376 & 1.402 & 0.445 & 0.412 & 1.172 & 1.359 & \textbf{1.584} \\
Sortino Ratio      & 2.147 & 2.158 & 0.722 & 0.702 & 1.811 & 2.070 & \textbf{2.317} \\
Information Ratio  & -0.221 & -0.241 & -0.701 & -0.728 & -0.175 & \textbf{0.188} & -- \\
Max Drawdown       & \textbf{0.162} & 0.164 & 0.289 & 0.295 & 0.190 & 0.217 & 0.165 \\
Calmar Ratio       & 2.114 & 2.136 & 0.858 & 0.839 & 1.874 & 2.266 & \textbf{2.300} \\
Avg Return         & 0.215 & 0.221 & 0.088 & 0.078 & 0.218 & 0.214 & \textbf{0.240} \\
Volatility         & 0.170 & 0.168 & 0.273 & 0.279 & 0.201 & 0.219 & \textbf{0.167} \\
\hline
\end{tabular}
\label{tab:nifty200_markowitz_m20_updated}
\end{table}

\begin{table}[htbp]
\centering
\footnotesize
\caption{Robustness metrics of portfolio strategies based on NIFTY 200 constituent stocks for a portfolio size of 30, where portfolio weights are obtained using the Markowitz optimization framework.}
\begin{tabular}{lccccccc}
\hline
\textbf{Metric} & $\boldsymbol{\pi}^{FULL}_p$ & $\boldsymbol{\pi}^{STR}_p$ & $\boldsymbol{\pi}^{FULL}_c$ & $\boldsymbol{\pi}^{STR}_c$ & $\boldsymbol{\pi}^{HSR}$ & $\boldsymbol{\pi}^{RND}$ & \textit{$\boldsymbol{\pi}^{MKT}$} \\
\hline
Sharpe Ratio       & 1.448 & 1.421 & 0.618 & 0.553 & 1.272 & 1.365 & \textbf{1.584} \\
Sortino Ratio      & 2.223 & 2.154 & 0.958 & 0.878 & 1.945 & 2.072 & \textbf{2.317} \\
Information Ratio  & -0.208 & -0.346 & -0.581 & -0.621 & -0.158 & \textbf{0.083} & -- \\
Max Drawdown       & \textbf{0.162} & 0.165 & 0.260 & 0.268 & 0.181 & 0.191 & 0.165 \\
Calmar Ratio       & 2.159 & 2.079 & 1.236 & 1.003 & 1.985 & 2.188 & \textbf{2.300} \\
Avg Return         & 0.219 & 0.214 & 0.133 & 0.113 & 0.222 & 0.208 & \textbf{0.240} \\
Volatility         & 0.166 & \textbf{0.165} & 0.259 & 0.264 & 0.189 & 0.201 & 0.167 \\
\hline
\end{tabular}
\label{tab:nifty200_markowitz_m30}
\end{table}

\begin{table}[htbp]
\centering
\footnotesize
\caption{Robustness metrics of portfolio strategies based on NIFTY 200 constituent stocks for a portfolio size of 40, where portfolio weights are obtained using the Markowitz optimization framework.}
\begin{tabular}{lccccccc}
\hline
\textbf{Metric} & $\boldsymbol{\pi}^{FULL}_p$ & $\boldsymbol{\pi}^{STR}_p$ & $\boldsymbol{\pi}^{FULL}_c$ & $\boldsymbol{\pi}^{STR}_c$ & $\boldsymbol{\pi}^{HSR}$ & $\boldsymbol{\pi}^{RND}$ & \textit{$\boldsymbol{\pi}^{MKT}$} \\
\hline
Sharpe Ratio       & 1.476 & 1.491 & 0.848 & 0.826 & 1.327 & 1.470 & \textbf{1.584} \\
Sortino Ratio      & 2.222 & 2.228 & 1.276 & 1.246 & 2.016 & 2.265 & \textbf{2.317} \\
Information Ratio  & -0.249 & -0.336 & -0.367 & -0.339 & -0.157 & \textbf{0.151} & -- \\
Max Drawdown       & \textbf{0.164} & 0.166 & 0.224 & 0.251 & 0.177 & 0.175 & 0.165 \\
Calmar Ratio       & 2.188 & 2.139 & 1.642 & 1.527 & 2.024 & \textbf{2.309} & 2.300 \\
Avg Return         & 0.221 & 0.220 & 0.188 & 0.184 & 0.223 & \textbf{0.259} & 0.240 \\
Volatility         & 0.165 & \textbf{0.164} & 0.238 & 0.255 & 0.183 & 0.190 & 0.167 \\
\hline
\end{tabular}
\label{tab:nifty200_markowitz_m40}
\end{table}

\begin{table}[htbp]
\centering
\footnotesize
\caption{Robustness metrics of portfolio strategies based on NIFTY 500 constituent stocks for a portfolio size of 10, where portfolio weights are obtained using the Markowitz optimization framework.}
\begin{tabular}{lccccccc}
\hline
\textbf{Metric} & $\boldsymbol{\pi}^{FULL}_p$ & $\boldsymbol{\pi}^{STR}_p$ & $\boldsymbol{\pi}^{FULL}_c$ & $\boldsymbol{\pi}^{STR}_c$ & $\boldsymbol{\pi}^{HSR}$ & $\boldsymbol{\pi}^{RND}$ & \textit{$\boldsymbol{\pi}^{MKT}$} \\
\hline
Sharpe Ratio       & 1.230 & 1.140 & 0.319 & 0.325 & 0.930 & 1.005 & \textbf{1.622} \\
Sortino Ratio      & 1.960 & 1.901 & 0.560 & 0.573 & 1.506 & 1.679 & \textbf{2.385} \\
Information Ratio  & \textbf{0.027} & -0.193 & -0.382 & -0.457 & -0.058 & 0.112 & -- \\
Max Drawdown       & \textbf{0.174} & 0.176 & 0.417 & 0.419 & 0.246 & 0.276 & \textbf{0.174} \\
Calmar Ratio       & 1.973 & 1.657 & 0.696 & 0.693 & 1.626 & 1.773 & \textbf{2.395} \\
Avg Return         & 0.225 & 0.203 & 0.074 & 0.082 & 0.211 & \textbf{0.250} & 0.229 \\
Volatility         & 0.189 & 0.184 & 0.364 & 0.368 & 0.242 & 0.271 & \textbf{0.159} \\
\hline
\end{tabular}
\label{tab:nifty500_markowitz_m10}
\end{table}

\begin{table}[htbp]
\centering
\footnotesize
\caption{Robustness metrics of portfolio strategies based on NIFTY 500 constituent stocks for a portfolio size of 20, where portfolio weights are obtained using the Markowitz optimization framework.}
\begin{tabular}{lccccccc}
\hline
\textbf{Metric} & $\boldsymbol{\pi}^{FULL}_p$ & $\boldsymbol{\pi}^{STR}_p$ & $\boldsymbol{\pi}^{FULL}_c$ & $\boldsymbol{\pi}^{STR}_c$ & $\boldsymbol{\pi}^{HSR}$ & $\boldsymbol{\pi}^{RND}$ & \textit{$\boldsymbol{\pi}^{MKT}$} \\
\hline
Sharpe Ratio       & 1.379 & 1.471 & 0.215 & 0.245 & 1.090 & 1.254 & \textbf{1.622} \\
Sortino Ratio      & 2.144 & 2.365 & 0.376 & 0.421 & 1.712 & 1.845 & \textbf{2.385} \\
Information Ratio  & -0.039 & 0.099 & -0.677 & -0.637 & -0.071 & \textbf{0.147} & -- \\
Max Drawdown       & 0.169 & \textbf{0.167} & 0.378 & 0.367 & 0.217 & 0.213 & 0.174 \\
Calmar Ratio       & 2.061 & 2.145 & 0.591 & 0.645 & 1.804 & 1.864 & \textbf{2.395} \\
Avg Return         & 0.218 & 0.240 & 0.050 & 0.060 & 0.214 & \textbf{0.256} & 0.229 \\
Volatility         & 0.170 & 0.170 & 0.331 & 0.329 & 0.211 & 0.214 & \textbf{0.159} \\
\hline
\end{tabular}
\label{tab:nifty500_markowitz_m20}
\end{table}

\begin{table}[htbp]
\centering
\footnotesize
\caption{Robustness metrics of portfolio strategies based on NIFTY 500 constituent stocks for a portfolio size of 30, where portfolio weights are obtained using the Markowitz optimization framework.}
\begin{tabular}{lccccccc}
\hline
\textbf{Metric} & $\boldsymbol{\pi}^{FULL}_p$ & $\boldsymbol{\pi}^{STR}_p$ & $\boldsymbol{\pi}^{FULL}_c$ & $\boldsymbol{\pi}^{STR}_c$ & $\boldsymbol{\pi}^{HSR}$ & $\boldsymbol{\pi}^{RND}$ & \textit{$\boldsymbol{\pi}^{MKT}$} \\
\hline
Sharpe Ratio       & 1.457 & 1.530 & 0.367 & 0.374 & 1.185 & 1.213 & \textbf{1.622} \\
Sortino Ratio      & 2.240 & \textbf{2.407} & 0.608 & 0.609 & 1.831 & 1.818 & 2.385 \\
Information Ratio  & -0.069 & \textbf{0.108} & -0.590 & -0.569 & -0.057 & -0.015 & -- \\
Max Drawdown       & \textbf{0.168} & 0.169 & 0.319 & 0.325 & 0.207 & 0.211 & 0.174 \\
Calmar Ratio       & 2.129 & 2.206 & 0.945 & 0.938 & 1.929 & 1.735 & \textbf{2.395} \\
Avg Return         & 0.219 & \textbf{0.239} & 0.091 & 0.086 & 0.217 & 0.225 & 0.229 \\
Volatility         & 0.162 & 0.163 & 0.299 & 0.301 & 0.198 & 0.200 & \textbf{0.159} \\
\hline
\end{tabular}
\label{tab:nifty500_markowitz_m30}
\end{table}

\begin{table}[htbp]
\centering
\footnotesize
\caption{Robustness metrics of portfolio strategies based on NIFTY 500 constituent stocks for a portfolio size of 40, where portfolio weights are obtained using the Markowitz optimization framework.}
\begin{tabular}{lccccccc}
\hline
\textbf{Metric} & $\boldsymbol{\pi}^{FULL}_p$ & $\boldsymbol{\pi}^{STR}_p$ & $\boldsymbol{\pi}^{FULL}_c$ & $\boldsymbol{\pi}^{STR}_c$ & $\boldsymbol{\pi}^{HSR}$ & $\boldsymbol{\pi}^{RND}$ & \textit{$\boldsymbol{\pi}^{MKT}$} \\
\hline
Sharpe Ratio       & 1.480 & \textbf{1.587} & 0.498 & 0.520 & 1.242 & 1.262 & 1.622 \\
Sortino Ratio      & 2.267 & \textbf{2.483} & 0.803 & 0.853 & 1.898 & 1.848 & 2.385 \\
Information Ratio  & -0.095 & \textbf{0.080} & -0.499 & -0.427 & -0.059 & -0.084 & -- \\
Max Drawdown       & 0.169 & 0.170 & 0.286 & 0.293 & 0.199 & 0.190 & \textbf{0.174} \\
Calmar Ratio       & 2.176 & \textbf{2.291} & 1.176 & 1.165 & 2.005 & 1.866 & 2.395 \\
Avg Return         & 0.217 & \textbf{0.236} & 0.121 & 0.128 & 0.216 & 0.215 & 0.229 \\
Volatility         & 0.160 & \textbf{0.158} & 0.279 & 0.285 & 0.190 & 0.185 & 0.159 \\
\hline
\end{tabular}
\label{tab:nifty500_m40_markowitz}
\end{table}

\begin{table}[htbp]
\centering
\footnotesize
\caption{Robustness metrics of portfolio strategies based on S\&P 500 constituent stocks for a portfolio size of 10, where portfolio weights are obtained using the Markowitz optimization framework.}
\begin{tabular}{lccccccc}
\hline
\textbf{Metric} & $\boldsymbol{\pi}^{FULL}_p$ & $\boldsymbol{\pi}^{STR}_p$ & $\boldsymbol{\pi}^{FULL}_c$ & $\boldsymbol{\pi}^{STR}_c$ & $\boldsymbol{\pi}^{HSR}$ & $\boldsymbol{\pi}^{RND}$ & \textit{$\boldsymbol{\pi}^{MKT}$} \\
\hline
Sharpe Ratio       & 0.814 & \textbf{0.851} & 0.748 & 0.737 & 0.766 & 0.808 & 0.899 \\
Sortino Ratio      & 1.226 & 1.209 & 1.052 & 1.035 & 1.124 & 1.203 & \textbf{1.247} \\
Information Ratio  & 0.037 & \textbf{0.097} & -0.081 & -0.090 & -0.067 & 0.069 & -- \\
Max Drawdown       & 0.197 & \textbf{0.198} & 0.168 & 0.169 & 0.186 & 0.158 & 0.175 \\
Calmar Ratio       & 1.271 & 1.247 & 1.219 & 1.196 & 1.230 & 1.283 & \textbf{1.310} \\
Avg Return         & 0.143 & \textbf{0.148} & 0.135 & 0.132 & 0.126 & 0.138 & 0.134 \\
Volatility         & 0.198 & 0.195 & 0.198 & 0.198 & 0.199 & 0.188 & \textbf{0.174} \\
\hline
\end{tabular}
\label{tab:markowitz_robustness_metrics}
\end{table}

\begin{table}[htbp]
\centering
\footnotesize
\caption{Robustness metrics of portfolio strategies based on S\&P 500 constituent stocks for a portfolio size of 20, where portfolio weights are obtained using the Markowitz optimization framework.}
\begin{tabular}{lccccccc}
\hline
\textbf{Metric} & $\boldsymbol{\pi}^{FULL}_p$ & $\boldsymbol{\pi}^{STR}_p$ & $\boldsymbol{\pi}^{FULL}_c$ & $\boldsymbol{\pi}^{STR}_c$ & $\boldsymbol{\pi}^{HSR}$ & $\boldsymbol{\pi}^{RND}$ & \textit{$\boldsymbol{\pi}^{MKT}$} \\
\hline
Sharpe Ratio       & 0.820 & 0.844 & 0.788 & 0.800 & 0.804 & 0.815 & \textbf{0.899} \\
Sortino Ratio      & 1.195 & 1.226 & 1.125 & 1.139 & 1.170 & 1.202 & \textbf{1.247} \\
Information Ratio  & -0.057 & \textbf{0.074} & -0.026 & -0.012 & -0.072 & -0.062 & -- \\
Max Drawdown       & 0.183 & 0.192 & 0.167 & 0.166 & 0.178 & 0.161 & \textbf{0.175} \\
Calmar Ratio       & 1.207 & 1.170 & 1.184 & 1.209 & 1.278 & 1.294 & \textbf{1.310} \\
Avg Return         & 0.131 & 0.140 & 0.140 & \textbf{0.142} & 0.127 & 0.132 & 0.134 \\
Volatility         & 0.182 & 0.186 & 0.191 & 0.191 & 0.189 & 0.182 & \textbf{0.174} \\
\hline
\end{tabular}
\label{tab:markowitz_robustness_metrics_20}
\end{table}

\begin{table}[htbp]
\centering
\footnotesize
\caption{Robustness metrics of portfolio strategies based on S\&P 500 constituent stocks for a portfolio size of 30, where portfolio weights are obtained using the Markowitz optimization framework.}
\begin{tabular}{lccccccc}
\hline
\textbf{Metric} & $\boldsymbol{\pi}^{FULL}_p$ & $\boldsymbol{\pi}^{STR}_p$ & $\boldsymbol{\pi}^{FULL}_c$ & $\boldsymbol{\pi}^{STR}_c$ & $\boldsymbol{\pi}^{HSR}$ & $\boldsymbol{\pi}^{RND}$ & \textit{$\boldsymbol{\pi}^{MKT}$} \\
\hline
Sharpe Ratio       & 0.853 & 0.828 & 0.792 & 0.791 & 0.804 & 0.810 & \textbf{0.899} \\
Sortino Ratio      & 1.245 & 1.198 & 1.138 & 1.139 & 1.164 & 1.170 & \textbf{1.247} \\
Information Ratio  & -0.046 & -0.001 & -0.028 & -0.031 & -0.100 & \textbf{0.123} & -- \\
Max Drawdown       & 0.178 & 0.188 & \textbf{0.162} & 0.173 & 0.176 & 0.172 & 0.175 \\
Calmar Ratio       & 1.210 & 1.166 & 1.217 & 1.225 & 1.256 & 1.209 & \textbf{1.310} \\
Avg Return         & 0.132 & 0.132 & \textbf{0.137} & 0.137 & 0.124 & 0.128 & 0.134 \\
Volatility         & 0.177 & 0.182 & 0.190 & 0.190 & 0.186 & 0.182 & \textbf{0.174} \\
\hline
\end{tabular}
\label{tab:markowitz_robustness_metrics_30}
\end{table}

\begin{table}[htbp]
\centering
\footnotesize
\caption{Robustness metrics of portfolio strategies based on S\&P 500 constituent stocks for a portfolio size of 40, where portfolio weights are obtained using the Markowitz optimization framework.}
\begin{tabular}{lccccccc}
\hline
\textbf{Metric} & $\boldsymbol{\pi}^{FULL}_p$ & $\boldsymbol{\pi}^{STR}_p$ & $\boldsymbol{\pi}^{FULL}_c$ & $\boldsymbol{\pi}^{STR}_c$ & $\boldsymbol{\pi}^{HSR}$ & $\boldsymbol{\pi}^{RND}$ & \textit{$\boldsymbol{\pi}^{MKT}$} \\
\hline
Sharpe Ratio       & 0.817 & 0.827 & \textbf{0.856} & 0.842 & 0.826 & 0.833 & 0.899 \\
Sortino Ratio      & 1.180 & 1.182 & \textbf{1.213} & 1.187 & 1.191 & 1.219 & 1.247 \\
Information Ratio  & -0.173 & -0.040 & \textbf{0.026} & 0.020 & -0.090 & 0.000 & -- \\
Max Drawdown       & 0.177 & 0.186 & \textbf{0.166} & 0.168 & 0.174 & 0.169 & 0.175 \\
Calmar Ratio       & 1.154 & 1.188 & \textbf{1.299} & 1.272 & 1.288 & 1.257 & 1.310 \\
Avg Return         & 0.123 & 0.129 & \textbf{0.144} & 0.143 & 0.127 & 0.143 & 0.134 \\
Volatility         & \textbf{0.174} & 0.180 & 0.185 & 0.187 & 0.183 & 0.183 & \textbf{0.174} \\
\hline
\end{tabular}
\label{tab:markowitz_robustness_metrics_40}
\end{table}

\clearpage

\section{Markowitz results}

\begin{table}[htbp]
\centering
\footnotesize
\caption{Performance comparison of Markowitz-weighted portfolio strategies in terms of signal-to-noise ratio, average returns, and volatility for portfolio sizes $M = \{10, 20, 30, 40\}$ over a holding horizon of $T = 125$ days, using NIFTY 200 stocks. Clearly, on average, the proposed investment strategy (\(\pi^{STR}_p\))  outperforms all other strategies in terms of the signal-to-noise ratio and average return while remaining competitive with the market portfolio (\(\pi^{MKT}\)).}
\label{tab:nifty_200_markowitz}
\begin{tabular*}{0.75\textwidth}{@{\extracolsep{\fill}} lcccccc|c }
\toprule
\textbf{M} &
\(\boldsymbol{\pi}^{FULL}_p\) &
\(\boldsymbol{\pi}^{STR}_p\) &
\(\boldsymbol{\pi}^{FULL}_c\) &
\(\boldsymbol{\pi}^{STR}_c\) &
\(\boldsymbol{\pi}^{HSR}\) &
\(\boldsymbol{\pi}^{RND}\) &
\textit{\(\boldsymbol{\pi}^{MKT}\)} \\
\midrule
\multicolumn{8}{c}{\scriptsize \textit{Signal-to-noise ratio}} \\
\midrule
10 & 0.9226 & 0.6956 & 0.1484 & 0.1541 & 0.6603 & 0.8955 & \textit{1.1905} \\
20 & 0.9559 & \textbf{0.9994} & 0.1705 & 0.1262 & 0.7671 & 0.7249 & \textit{1.1905} \\
30 & 1.0449 & \textbf{1.0564} & 0.2361 & 0.2190 & 0.8488 & 0.8316 & \textit{1.1905} \\
40 & 1.1401 & \textbf{1.2129} & 0.4932 & 0.4080 & 0.8854 & 0.9958 & \textit{1.1905} \\
\midrule
\multicolumn{8}{c}{\scriptsize \textit{Average Return}} \\
\midrule
10 & 0.1063 & 0.0973 & 0.0346 & 0.0361 & 0.1059 & 0.1281 & \textit{0.1238} \\
20 & 0.1094 & \textbf{0.1165} & 0.0325 & 0.0245 & 0.1084 & 0.1107 & \textit{0.1238} \\
30 & 0.1145 & \textbf{0.1155} & 0.0496 & 0.0391 & 0.1115 & 0.1119 & \textit{0.1238} \\
40 & 0.1183 & \textbf{0.1201} & 0.0824 & 0.0748 & 0.1119 & 0.1210 & \textit{0.1238} \\
\midrule
\multicolumn{8}{c}{\scriptsize \textit{Volatility}} \\
\midrule
10 & 0.1152 & 0.1399 & 0.2333 & 0.2343 & 0.1604 & 0.1431 & \textit{0.1040} \\
20 & 0.1144 & 0.1166 & 0.1910 & 0.1945 & 0.1413 & 0.1527 & \textit{0.1040} \\
30 & 0.1096 & 0.1093 & 0.2102 & 0.1785 & 0.1313 & 0.1346 & \textit{0.1040} \\
40 & 0.1038 & \textbf{0.0991} & 0.1671 & 0.1832 & 0.1263 & 0.1215 & \textit{0.1040} \\
\bottomrule
\end{tabular*}
\end{table}


\begin{table}[htbp]
\centering
\footnotesize
\caption{Performance comparison of Markowitz-weighted portfolio strategies in terms of signal-to-noise ratio, average returns, and volatility for portfolio sizes $M = \{10, 20, 30, 40\}$ over a holding horizon of $T = 125$ days, using NIFTY 500 stocks. Clearly, on average, the proposed investment strategy (\(\pi^{STR}_p\))  outperforms all other strategies in terms of the signal-to-noise ratio and average return while remaining competitive with the market portfolio (\(\pi^{MKT}\)).}
\label{tab:nifty_500_markowitz}
\begin{tabular*}{0.75\textwidth}{@{\extracolsep{\fill}} lcccccc|c }
\toprule
\textbf{M} &
\(\boldsymbol{\pi}^{FULL}_p\) &
\(\boldsymbol{\pi}^{STR}_p\) &
\(\boldsymbol{\pi}^{FULL}_c\) &
\(\boldsymbol{\pi}^{STR}_c\) &
\(\boldsymbol{\pi}^{HSR}\) &
\(\boldsymbol{\pi}^{RND}\) &
\textit{\(\boldsymbol{\pi}^{MKT}\)} \\
\midrule
\multicolumn{8}{c}{\scriptsize \textit{Signal-to-noise ratio}} \\
\midrule
10 & 0.6707 & 0.6232 & 0.0834 & 0.1037 & 0.5339 & 0.5128 & \textit{0.8769} \\
20 & 0.7647 & 0.7044 & 0.0144 & 0.0458 & 0.6253 & 0.7647 & \textit{0.8769} \\
30 & 0.7443 & \textbf{0.7717} & 0.1294 & 0.1152 & 0.6780 & 0.7249 & \textit{0.8769} \\
40 & 0.7761 & \textbf{0.7776} & 0.2623 & 0.2511 & 0.7218 & 0.7613 & \textit{0.8769} \\
\midrule
\multicolumn{8}{c}{\scriptsize \textit{Average Return}} \\
\midrule
10 & 0.1010 & \textbf{0.1050} & 0.0224 & 0.0271 & 0.1039 & 0.1050 & \textit{0.1249} \\
20 & 0.1116 & \textbf{0.1181} & 0.0032 & 0.0099 & 0.1121 & 0.1232 & \textit{0.1249} \\
30 & 0.1122 & \textbf{0.1257} & 0.0278 & 0.0245 & 0.1153 & 0.1083 & \textit{0.1249} \\
40 & 0.1143 & \textbf{0.1240} & 0.0571 & 0.0555 & 0.1150 & 0.1046 & \textit{0.1249} \\
\midrule
\multicolumn{8}{c}{\scriptsize \textit{Volatility}} \\
\midrule
10 & 0.1506 & 0.1684 & 0.2692 & 0.2965 & 0.2022 & 0.2048 & \textit{0.1424} \\
20 & 0.1559 & 0.1677 & 0.2247 & 0.2158 & 0.1793 & 0.1611 & \textit{0.1424} \\
30 & 0.1507 & 0.1629 & 0.2151 & 0.2124 & 0.1701 & 0.1494 & \textit{0.1424} \\
40 & 0.1472 & 0.1595 & 0.2178 & 0.2208 & 0.1593 & 0.1374 & \textit{0.1424} \\
\bottomrule
\end{tabular*}
\end{table}

\begin{table}[htbp]
\centering
\footnotesize
\caption{Performance comparison of Markowitz-weighted portfolio strategies in terms of signal-to-noise ratio, average returns, and volatility for portfolio sizes $M = \{10, 20, 30, 40\}$ over a holding horizon of $T = 125$ days, using S\&P 500 stocks. Clearly, on average, the proposed investment strategy (\(\pi^{STR}_p\))  outperforms all other strategies in terms of average return, while remaining competitive with the market portfolio and other performance measures.}
\label{tab:sp_500_markowitz}
\begin{tabular*}{0.75\textwidth}{@{\extracolsep{\fill}} lcccccc|c }
\toprule
\textbf{M} &
\(\boldsymbol{\pi}^{FULL}_p\) &
\(\boldsymbol{\pi}^{STR}_p\) &
\(\boldsymbol{\pi}^{FULL}_c\) &
\(\boldsymbol{\pi}^{STR}_c\) &
\(\boldsymbol{\pi}^{HSR}\) &
\(\boldsymbol{\pi}^{RND}\) &
\textit{\(\boldsymbol{\pi}^{MKT}\)} \\
\midrule
\multicolumn{8}{c}{\scriptsize \textit{Signal-to-noise ratio}} \\
\midrule
10 & 0.5089 & 0.6507 & 0.6502 & 0.6774 & 0.4549 & 0.5855 & \textit{0.6728} \\
20 & 0.5766 & \textbf{0.6791} & 0.7597 & 0.7681 & 0.4859 & 0.4736 & \textit{0.6728} \\
30 & 0.6212 & 0.5927 & 0.6518 & 0.7775 & 0.4872 & 0.4746 & \textit{0.6728} \\
40 & 0.8504 & 0.7907 & 0.7713 & 0.7684 & 0.7569 & 0.8484 & \textit{0.8186} \\
\midrule
\multicolumn{8}{c}{\scriptsize \textit{Average Return}} \\
\midrule
10 & 0.0575 & \textbf{0.0741} & 0.0623 & 0.0607 & 0.0529 & 0.0724 & \textit{0.0636} \\
20 & 0.0542 & \textbf{0.0717} & 0.0668 & 0.0669 & 0.0532 & 0.0591 & \textit{0.0636} \\
30 & 0.0578 & \textbf{0.0626} & 0.0600 & 0.0597 & 0.0524 & 0.0528 & \textit{0.0636} \\
40 & 0.0558 & 0.0595 & 0.0619 & 0.0597 & 0.0546 & 0.0632 & \textit{0.0636} \\
\midrule
\multicolumn{8}{c}{\scriptsize \textit{Volatility}} \\
\midrule
10 & 0.1129 & 0.1139 & 0.0957 & 0.0958 & 0.1163 & 0.1237 & \textit{0.0946} \\
20 & 0.0941 & 0.1055 & 0.0879 & 0.0871 & 0.1095 & 0.1248 & \textit{0.0946} \\
30 & 0.0931 & 0.1055 & 0.0920 & 0.0921 & 0.1076 & 0.1112 & \textit{0.0946} \\
40 & 0.0904 & 0.1050 & 0.0906 & 0.0921 & 0.1069 & 0.1149 & \textit{0.0946} \\
\bottomrule
\end{tabular*}
\end{table}

\clearpage
